\documentclass[prb,twocolumn,superscriptaddress,showpacs,floatfix]{revtex4}

\usepackage[english]{babel}
\usepackage[dvips]{graphicx}
\usepackage{amsmath}
\usepackage{amssymb}
\usepackage{latexsym}
\newcommand{\beq}{\begin{equation}}
\newcommand{\eeq}{\end{equation}}
\newcommand{\barr}{\begin{eqnarray}}
\newcommand{\earr}{\end{eqnarray}}
\newcommand{\ket}[1]{\left\vert#1\right\rangle}
\newcommand{\bra}[1]{\left\langle#1\right\vert}
\newcommand{\Ham}{\mathcal H}

\begin{document}

\title{Long time dynamics
following a quench in an integrable quantum spin chain: \\ 
local versus non-local operators and effective thermal behavior}

\author{Davide Rossini}
\altaffiliation{Present institution:
  {\it Scuola Normale Superiore,
  Piazza dei Cavalieri 7, I-56126 Pisa, Italy.}}
\affiliation{International School for Advanced Studies (SISSA),
  Via Beirut 2-4, I-34014 Trieste, Italy}

\author{Sei Suzuki}
\affiliation{Department of Physics and Mathematics, 
  Aoyama Gakuin University, Fuchinobe, Sagamihara 229-8558, Japan}

\author{Giuseppe Mussardo}
\affiliation{International School for Advanced Studies (SISSA),
  Via Beirut 2-4, I-34014 Trieste, Italy}
\affiliation{International Centre for Theoretical Physics (ICTP),
  I-34014 Trieste, Italy}
\affiliation{Istituto Nazionale di Fisica Nucleare, Sezione di Trieste}

\author{Giuseppe E. Santoro}
\affiliation{International School for Advanced Studies (SISSA),
  Via Beirut 2-4, I-34014 Trieste, Italy}
\affiliation{CNR-INFM Democritos National Simulation Center,
  Via Beirut 2-4, I-34014 Trieste, Italy}
\affiliation{International Centre for Theoretical Physics (ICTP),
  I-34014 Trieste, Italy}

\author{Alessandro Silva}
\affiliation{International Centre for Theoretical Physics (ICTP),
  I-34014 Trieste, Italy}

\date{\today}

\begin{abstract}

  We study the dynamics of the quantum Ising chain following a zero-temperature
  quench of the transverse field strength. Focusing on the behavior of two-point spin 
  correlation functions, we show that the correlators of 
  the order parameter display an effective asymptotic thermal behavior, 
  i.e., they decay exponentially to zero, with 
  a phase coherence rate and a correlation length dictated by the equilibrium law
  with an effective temperature set by the energy of the initial state.
  On the contrary, the two-point correlation functions of 
  the transverse magnetization or the density-of-kinks operator decay as a power-law 
  and do not exhibit thermal behavior.
  We argue that the different behavior is linked to the locality of the corresponding 
  operator with respect to the quasi-particles of the model: non-local operators, 
  such as the order parameter, behave thermally, while local ones do not. 
  We study which features of the two-point correlators are a consequence of the
  integrability of the model by analizing their robustness with respect to
  a sufficiently strong integrability-breaking term.

\end{abstract}



\maketitle

\section{Introduction} \label{sec:intro}

The study of the non-equilibrium dynamics of strongly correlated
quantum many body systems has lately received an increasing amount of attention.
While the first theoretical studies in this area were performed
a few decades ago~\cite{niemeijer67,barouch69,mccoy1,mccoy2}, interest
on this subject has been confined for a long time to the theoretical literature.
The main trigger behind the recent experimental advances on this subject
has been the impressive progress in manipulating cold atomic gases,
which resulted in the first experiments exploring the coherent
non-equilibrium dynamics of strongly interacting 
quantum systems in a controllable way~\cite{greiner02,sadler06,weiler08,kinoshita06}.

On the practical side, the main advantage of studying many-body physics with cold atoms
is the detailed microscopic knowledge of the Hamiltonian describing these systems.
This fact, together with the possibility of controlling the system's parameters
with high accuracy and with the availability of long coherence times, has made it
possible to study the unitary dynamics of many-body systems~\cite{lewenstein07,bloch08}, 
such as Hubbard-like Hamiltonians~\cite{jaksch98} or artificial spin chain
models~\cite{duan03,jane03}, a topic that can hardly be addressed
in traditional solid-state environments.
Coherent non-equilibrium dynamics has been brilliantly demonstrated by the observation
of collapse and revival cycles in systems quenched across a superfluid-to-Mott insulator
quantum phase transition~\cite{greiner02} and in the study of the formation
of topological defects during a quench of trapped atomic gases
through a critical point~\cite{sadler06,weiler08}.

One of the most fundamental topics in quantum statistical mechanics which is presently
being studied in this context is the connection between ergodicity,
non-integrability, and thermalization in the dynamics of strongly interacting (but isolated)
many-body systems. An important experiment addressing this issue was recently
reported in Ref.~[\onlinecite{kinoshita06}], where the absence of thermalization
in the coherent non-equilibrium evolution of a closely integrable one-dimensional
Bose gas was observed. Motivated by these experimental findings,
an increasing number of theoretical studies focusing on the dynamics following
a sudden perturbation (a {\it quantum quench}) have been 
performed~\cite{igloi00,sengupta04,cazalilla06,calabrese06,rigol07,rigol08,rigol09,fiorettomussardo,kollath07,roux09,manmana07,eckstein09,cramer08,barthel08,gangardt08,kollar08,kollar08b,paper1,calabrese09,sotiriadis09,iucci09,heisquench09,moeckel08}. 

Integrability is believed to play a crucial role in the relaxation process:
in analogy to the well-known Fermi-Pasta-Ulam scenario in classical systems,
integrable systems are not expected to thermalize, but to be sensitive to the
specifics of the initial state~\cite{rigol07,rigol08,rigol09,fiorettomussardo}.
This understanding was distilled into the proposal of a generalized Gibbs ensemble,
keeping track of the initial value of all the constants of motion~\cite{rigol07}
and constructed to describe the steady state reached after a quench.
Several works have tested the conditions of applicability of such Gibbs distribution
and its drawbacks~\cite{cazalilla06,calabrese06,rigol07,fiorettomussardo,gangardt08,kollar08,kollar08b,iucci09}.
In particular, for a special quench in a 1D Bose-Hubbard model~\cite{cramer08}, 
for integrable systems with free quasiparticles~\cite{barthel08} 
and for the computation of {\it one-point} correlation functions for a specific class 
of quench processes in otherwise generic integrable systems~\cite{fiorettomussardo}, 
the long-time limit of the dynamics was shown to be well described by 
the generalized Gibbs ensemble. Moreover, the generalized Gibbs ensemble was
shown to correctly predict the asymptotic momentum distribution
functions for a variety of models and quenches~\cite{calabrese06,rigol07,kollar08,cazalilla06}. 
However, it should be pointed out that neglection of correlations of the occupation of different
quasi-particle modes generally leads to incorrect predictions for
the noise and higher order correlators~\cite{gangardt08}. 
Turning to non-integrable systems, in this case thermalization is expected
to occur in general: this has been numerically observed in some
circumstances~\cite{rigol08,rigol09,kollath07,roux09,manmana07,eckstein09},
while the transition from integrable to non-integrable has been shown, at least in 
small lattices of interacting bosonic or fermionic particles~\cite{rigol09},
to take the form of a crossover. 
The mechanism of thermalization has been conjectured and numerically tested
for certain systems~\cite{rigol08,rigol09} to be analogous to the one
proposed by Deutsch~\cite{deutsch91} and Srednicki~\cite{srednicki94}
for systems with a classically chaotic counterpart.
Nonetheless, it is worth pointing out that in some specific cases, 
like for gapped systems, the validity of this scenario has been questioned
(the points raised include finite-size effects~\cite{roux09} and the importance
of rare events~\cite{biroli09}, which may drive the behavior of the long-time dynamics), 
and the problem is still under debate.

\begin{figure}[!t]
  \begin{center}
    \includegraphics[scale=0.33]{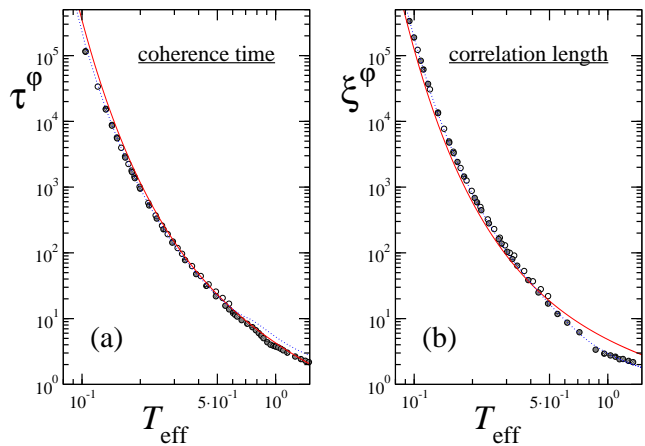}
    \caption{(color online). Phase coherence time (a)
      and correlation length (b) extracted from the asymptotic decay
      of the two-point order parameter correlations,
      as a function of the effective temperature.
      Data are for quenches ending in the ferromagnetic phase of the Ising chain
      at $\Gamma = 0.5$ [see Eq.~\eqref{eq:model}].
      Symbols refer to the case of a quench and correspond to different values
      of $\Gamma_0$ (empty symbols are for $\Gamma_0 < \Gamma$,
      while filled ones are for $\Gamma_0 > \Gamma$);
      continuous red curves denote the equilibrium values
      at finite temperatures, while dotted blue curves correspond to
      values obtained with a semiclassical analysis generalized to
      non-equilibrium cases.}
    \label{fig:Intro_thermal}
  \end{center}
\end{figure}

The purpose of this work is to go one step beyond the scenario proposed above
and show that in quantum many-body systems the presence or absence of thermal
behavior after a quantum quench does not exclusively depend on the integrability
of the model, but also on the considered {\em observable}.
In particular, we argue that, even in a completely integrable system, 
the asymptotics of the {\em two-point} correlation functions of an observable
which is non-local with respect to the quasi-particle fields displays
thermal behavior, while this is definitely not the case for a local one.
This scenario may be violated in some specific and isolated cases, such as 
for the transverse-field correlations or the density-of-kinks operator
after a quench towards the critical point.
In this context, local and non-local operators refer to the 
structure of their matrix elements on the basis of quasi-particles: local means 
that the operator couples a {\em finite} number of states, while non-local means
that it couples {\em all} states. 
To show the different behavior of the correlation functions of local and non-local 
operators, we study quantum quenches in a quantum Ising chain, 
focusing our attention on the asymptotics of the two-point correlators of the order
parameter $\sigma^x$ and of the transverse magnetization $\sigma^z$ after a quench
of the transverse field. We explicitly show that the correlator of the order parameter,
which is non-locally expressed in terms of the fermions which diagonalize the model,
asymptotically displays a thermal-like behavior characterized by an exponential
decay both in time and in space. As we already reported in Ref.~[\onlinecite{paper1}],
the autocorrelation function of the order parameter after a quench takes the form
\begin{eqnarray}
\bra{\cal B} \sigma^x_{i}(t) \, \sigma^x_{i}(0) \ket{\cal B} \approx e^{-t/\tau^{\varphi}_Q},
\end{eqnarray}
with a phase coherence time $\tau^\varphi_Q$ depending only on the effective
temperature $T_{\rm eff}$ set by the ground state energy of the initial Hamiltonian 
$\ket{\cal B}$ (below also called {\em boundary state}), and on the energy gap $\Delta$ of the
final Hamiltonian. Here we are going to show that such thermal behavior
is not limited to the equal-site autocorrelation function, but also pertains 
to the large distance behavior of the equal-time correlator (asymptotically,
for large times $t$), which behaves as
\begin{eqnarray}
\bra{\cal B} \sigma^x_{i+r}(t) \, \sigma^x_{i}(t) \ket{\cal B} \approx e^{-r/\xi^{\varphi}_Q},
\end{eqnarray} 
with a correlation length $\xi^{\varphi}_Q$ again determined by $T_{\rm eff}$
and $\Delta$. Remarkably, as shown in Fig.~\ref{fig:Intro_thermal}(a), the
dependence of $\tau^\varphi_Q$ on $T_{\rm eff}$ (black circles) quantitatively
agrees with that of the corresponding phase coherence time $\tau^\varphi_T$ \cite{sachdev97}
at thermal equilibrium as a function of the temperature (continuous red line). 
A similar behavior is observed for the correlation length
[see Fig.~\ref{fig:Intro_thermal}(b)].
All of these results can be understood in terms of the generalization to the non-equilibrium case
of a semiclassical analysis originally developed for thermal equilibrium~\cite{sachdev97}, 
which explicitly takes into account the effect of 
quasiparticles generated after the quench [see, e.g., the dotted blue line 
in Fig.~\ref{fig:Intro_thermal}(a)-(b)]. 
This thermal behavior is not present at all in the asymptotics of the correlator of
the transverse magnetization or in the nearest-neighbor spatial correlations of the
longitudinal magnetization (i.e., representing the density of kinks), that are operators 
which turn out to be local in the fermionic quasiparticles of the Ising chain. 

The fact that quasiparticles play a fundamental role in the physics of thermalization
is not at all surprising: the essence of quantum integrability is the possibility
to describe a system in terms of quasiparticles subject to purely elastic
and factorizable scattering. A non-integrable system, in this sense,
is just characterized by the absence of well defined quasiparticles:
the scattering of excitations generically leads to a cascade 
redistributing the initial energy among the low-energy degrees of freedom.
The results discussed in this paper indicate that if an observable is sensitive
to the detailed scattering properties of such degrees of freedom, i.e., it is local,
then its asymptotics will be affected by the presence or absence of integrability.
In turn, if such relation is non-local, then, as for the equilibrium physics,
we expect that for its multi-point correlation functions only some gross
low-energy features of the model are relevant, and determined by the universality class.
We stress again that here locality is defined {\it with respect to the quasiparticles},
while the original degrees of freedom do not play a major role.

The paper is organized as follows: in Sec.~\ref{sec:model} we define our model
and the techniques used to study its dynamics after a quench.
In order to clarify the meaning of locality in this context, 
we also discuss the continuum formulation of the model
in terms of free Majorana fermions, and enlighten the physical content of the initial
state $\ket{\cal B}$ and its connection with boundary states in statistical field 
theory~\cite{ghoshalzam}.
In Sec.~\ref{sec:boundary} we summarize the main formulas for studying the time evolution of the system.
In Sec.~\ref{sec:XXquench} we discuss the order parameter correlation functions: 
we consider in details the behavior after a quench
showing that it is asymptotically thermal, both in time and in space. 
In Sec.~\ref{sec:nonthermal} we focus on two examples of local operators
(namely, the transverse field correlations and the density of kinks,
which is essentially the nearest-neighbor space correlator of the order parameter),
which, on the contrary, do not exhibit thermalization. 
The non-thermal behavior can be nonetheless described by means of 
a generalized Gibbs ensemble, as discussed in Sec.~\ref{subsec:Gibbs}.
A brief discussion of the effects of integrability breaking
is sketched in Sec.~\ref{sec:NonInt}, where we provide numerical data showing
the gradual disappearance of a typical non-thermal feature in the asymptotic 
spatial behavior of the order parameter correlators.
Finally, in Sec.~\ref{sec:concl} we draw our conclusions.

\section{The Model: Generalities, continuum limit and boundary state} \label{sec:model}

In this paper we study the spin-$1/2$ quantum Ising chain in a transverse
magnetic field~\cite{sachdev}, which is characterized by the Hamiltonian
\beq
   \Ham(\Gamma) = -J \sum_{j=1}^{L} \left[ \sigma^x_j \sigma^x_{j+1}
   + \Gamma \sigma^z_j \right] \;,
   \label{eq:model}
\eeq
where $L$ is the number of spins in the chain and $\sigma_j^\alpha$ ($\alpha = x,y,z$) 
are the Pauli matrices relative to the $j$th spin. Hereafter we impose periodic 
boundary conditions, and use a system of units $J=1$, $\hbar=1$, and $k_B=1$. 
The quantum Ising chain is the prototype of an exactly solvable quantum 
system and is characterized by two mutually dual gapped phases, a quantum 
paramagnet ($\Gamma>1$) and a ferromagnet ($\Gamma<1$), separated by 
a quantum critical point at $\Gamma_c = 1$.

In the following we will consider sudden quenches of the transverse magnetic 
field strength $\Gamma$: after initializing the system in the 
ground state $\ket{\phi(\Gamma_0)} \equiv \ket{\cal B}$ of the 
Hamiltonian $\Ham(\Gamma_0)$, the field strength is suddenly changed 
at time $t=0$ from $\Gamma_0$ to a new value $\Gamma \neq \Gamma_0$. 
Consequently, the state will evolve according to the new Hamiltonian $\Ham(\Gamma)$:
\beq
   \ket{\phi_t} = e^{- i \Ham(\Gamma) t} \ket{\cal B} \;.
   \label{eq:psit}
\eeq
%

\subsection{Lattice formalism}

Before entering into the details of the non-equilibrium dynamics of this model,
let us briefly set the notation by reviewing the diagonalization~\cite{lieb61,pfeuty70}
of the Hamiltonian~\eqref{eq:model}. 
Introducing Jordan-Wigner fermions $c_j^\dagger$ defined by
\beq
\sigma^+_j \equiv c^{\dagger}_j  \exp \Big( i \pi \sum_{l=1}^{j-1} c^{\dagger}_l c_l \Big)
\label{stringfermion}
\eeq
and omitting constant terms, the Hamiltonian in Eq.~\eqref{eq:model} takes the quadratic form
\barr
\nonumber \Ham & = & -  \sum_{j=1}^{L-1}
\left[ c^\dagger_j \, c_{j+1} + c^\dagger_j c^\dagger_{j+1} + h.c. \right]
- 2 \Gamma \sum_{j=1}^L c^\dagger_j c_j  \\
{} & {} & + (-1)^{N_F} \left[ c^\dagger_L c_1 + c^\dagger_L c^\dagger_1 + h.c. \right] \;.
\label{eq:modelJW}
\earr
The last term originates from the periodic boundary conditions imposed to the spins and
its sign depends on the parity of the total number $N_F$ of $c$-fermions. 
Specifically, if $N_F$ is odd, then all the bonds are identical and periodic boundary 
conditions on the fermions are imposed ($c_{L+1} \equiv c_1$). 
Antiperiodic boundary conditions ($c_{L+1} \equiv - c_1$) are instead 
appropriate when $N_F$ is even.
The Hamiltonian in Eq.~\eqref{eq:modelJW} conserves the fermion parity,
therefore it can be formally split in two parts acting on different Fermion-parity
subspaces, even ($+$) and odd ($-$):
$\Ham = \Ham^+ + \Ham^-$, where $\Ham^{\pm} \equiv {\cal P}^{\pm} \Ham {\cal P}^{\pm}$
denote the even/odd subspaces, and ${\cal P}^{\pm}$ the associated projectors.
Since in the following we will consider initial 
states with an even number of fermions, we will focus our attention 
on the even sector only.

The diagonalization of the Hamiltonian now proceeds in momentum space.
Writing the $c$-fermions as $c_j = \frac{1}{\sqrt{L}} \sum_k e^{i k j} c_k$,
where $k$ is
\beq
\label{eq:abc}
k = \pm \frac{\pi (2 n +1)}{L} \qquad {\rm with} \quad n=0, \cdots, L/2-1 \;,
\eeq
the Fourier representation of Eq.~\eqref{eq:modelJW} in the even sector, $\Ham^+$,
becomes a sum of independent terms 
\beq
{\cal H}^+=\sum_{k>0} \bar{c}^{\dagger}_k\;{\bf H}_k\;\bar{c}_k \;,
\eeq
where we introduced the Nambu vector $\bar{c}^{\dagger}_k= (c^\dagger_k \;,\, c_{-k})$, and 
\beq
{\bf H}_k=\left(
\begin{array}{cc} a_k & - i b_k \\ i b_k & -a_k \end{array} \right)
\eeq
with $a_k = 2 (\Gamma - \cos k)$ and $b_k = 2 \sin k$.
This Hamiltonian is easily diagonalized through a Bogoliubov rotation,
defining the new fermionic variables
\beq
\bar{{\cal A}}_k={\bf R}_k(\Gamma)\;\bar{c}_k
\label{eq:nambudef}
\eeq
where $\bar{{\cal A}}_k^T=({\cal A}_k \; ,\,{\cal A}^{\dagger}_{-k})$.
The rotation ${\bf R}$ is explicitly given by 
\beq
{\bf R}_k(\Gamma)=\left(
\begin{array}{cc} u^*_k & v^*_k \\ -v_k & u_k \end{array} \right) \;.
\eeq
Here 
\beq
\begin{array}{c} \displaystyle
\displaystyle
u_k = \frac{\epsilon_k + a_k}{\sqrt{2 \epsilon_k (\epsilon_k + a_k)}} \,, \quad
v_k = \frac{i b_k}{\sqrt{2 \epsilon_k (\epsilon_k + a_k)}} \;,
\end{array}
\label{eq:enerdef}
\eeq
while $\epsilon_k = \sqrt{a_k^2 + b_k^2} = 2 \sqrt{ \Gamma^2  - 2 \Gamma \cos k + 1}$
is the dispersion of the quasiparticles in terms of which 
\beq
\Ham^+ = \sum_{k>0} \epsilon_k \big( {\cal A}^\dagger_k {\cal A}_k
+ {\cal A}^\dagger_{-k} {\cal A}_{-k} -1 \big) \;.
\label{eq:hamdiag}
\eeq
The ground state of the system is the vacuum of the Bogoliubov quasiparticles 
defined in Eq.~\eqref{eq:nambudef}. These become gapless at the quantum critical 
point $\Gamma_c$, where the gap $\Delta \equiv \epsilon_0 = 2 \vert 1-\Gamma \vert$ vanishes.

\subsection{Continuum formalism}

The continuum formulation of the quantum Ising model enlightens other properties,
such as the local/non-local nature of its operators, and is helpful
in understanding the properties of the initial state $\ket{\cal B}$
associated to the quench process.
Let us briefly discuss the main concepts of this formulation.

In the vicinity of the quantum critical point, taking the continuum limit
($a \rightarrow 0$, where $a$ is the lattice spacing), it is well known that
the Ising model becomes a field theory of free relativistic Majorana fermions
$\psi_+(r,t)$ and $\psi_-(r,t)$~\cite{mussardobook}.
These are real fields, $\psi_{\pm}^{\dagger}(r,t) = \psi_{\pm}(r,t)$,
and obey the equal-time anti-commutation relations 
\beq
\{\psi_i(r,t),\psi_j(r',t)\} \,=\ \delta_{ij}\,\delta(r-r') \;.  
\label{etanticom}
\eeq 
The fermion particle excitations satisfy a relativistic dispersion
$E(p) = \sqrt{ p^2 + \Delta^2}$ with a mass $m=\Delta$, and 
their dynamics is described by the Hamiltonian 
\beq 
{\cal H} \,=\, \int dr \left[i \psi_+ \partial_r \psi_+  - 
i \psi_- \partial_r\psi_-  - 2 i m \psi_+ \psi_- \right]\,\,\,.
\label{DiracHamiltonian}
\eeq
The mode expansion of the fermion fields is given by 
\beq
\psi_{\pm}(r,t) = \int_{-\infty}^{\infty} \frac{d k}{2\pi} 
\left[\alpha_\pm(k) {\cal A}(k) e^{-i E t + i k r} + h.c. \right]  
\label{firstmodeexpansion}
\eeq
where $\{{\cal A}(k),{\cal A}^{\dagger}(k')\} = 2 \pi \,\delta(k-k')$, and 
\beq 
\alpha_{\pm}(k) \,=\,\omega^{\pm 1} \,\sqrt{\frac{E \pm k}{ 2 E}} \;,
\label{alpha+-}
\eeq 
with $\omega = e^{i \pi/4}$.
In order to express later in a more concise form both the operators 
and the matrix elements of the model, it is convenient to 
re-write the mode expansion in a slightly different way. 
This can be done using the identity 
\[
\lim_{\delta\rightarrow 0} (\delta + i u)^{\pm \frac{1}{2}} = 
\theta(u) \left(\omega \sqrt{|u|}\right)^{\pm 1} + 
\theta(-u) \left(\omega^{-1} \sqrt{|u|}\right)^{\pm 1} 
\]
$\theta(u)$ being the Heaviside theta, so that Eq.~\eqref{firstmodeexpansion} becomes 
\beq
\psi_{\pm}(r,t) = \sqrt{\frac{\Delta}{2}}\,\int_{-\infty}^{+\infty}
\frac{du}{2\pi u} e^{-i E t + i p r} \,(\delta + i u)^{\pm \frac{1}{2} }\,
\hat\psi(u) \;,
\label{modeexpansionu}
\eeq
where $E = \Delta (u + u^{-1})/2$, $p = \Delta (u - u^{-1})/2$ and 
\beq
\hat\psi(u) = \frac{1}{\sqrt{2}}\left[\theta(u) {\cal A}(u)
- \theta(-u) {\cal A}^{\dagger}(-u)\right] \;. 
\eeq
The annihilation and creation operators, ${\cal A}(u)$ and ${\cal A}^{\dagger}(u)$, 
are the continuous version, in the variable $u$, of the lattice operators 
${\cal A}_k$ and ${\cal A}^{\dagger}_k$.  
The vacuum state $\ket{0}$ is identified by the conditions ${\cal A}(u) \ket{0} = 0$. 
Notice that $\hat\psi(u)$ is an annihilation operator for $u >0$ and a creation 
operator for $u < 0$. It is also useful to define 
\beq
\psi_0(r,t) = \sqrt{\frac{\Delta}{2}}\, \int_{-\infty}^{+\infty} \frac{du}{2\pi u} 
\hat\psi(u) \, e^{-i E t + i p r} \;.
\eeq 
Since the work of Fradkin and Susskind~\cite{FradkinSusskind}, it became well 
known that, in the continuum limit, the lattice operator $\sigma^x_r$ becomes 
\beq 
\sigma^x_r \rightarrow \left\{
\begin{array}{cc}
 \sigma(r,t) & \; {\rm for} \; \Gamma > 1 \\
 \mu(r,t)    & \; {\rm for} \; \Gamma < 1 
\end{array}
\right. \;, \label{order-disorder}
\eeq
where $\sigma(r,t)$ is the ``order parameter'',
while $\mu(r,t)$ is the ``disorder parameter''.
On the other hand, for the lattice operator $\sigma^z_r$ we have
\beq
\sigma^z_r \rightarrow \left\{
\begin{array}{cc}
 \epsilon(r,t) & \; {\rm for} \; \Gamma > 1 \\
-\epsilon(r,t) & \; {\rm for} \; \Gamma < 1 
\end{array}
\right. \;,
\label{energyoperator}
\eeq
where $\epsilon(r,t)$ is the ``energy operator''.
Eqs.~\eqref{order-disorder} and~\eqref{energyoperator} express
the self-duality of the Ising model. In particular, for $\Gamma > 1$,
$\sigma(r,t)$ is a $Z_2$ odd operator with non-zero fermion number
while $\mu(r,t)$ is a $Z_2$ even operator with zero fermion number: so, 
$\sigma(r,t)$ has non-zero matrix elements only on odd number of fermions,
while $\mu(r,t)$ has non-zero matrix elements only on even number of fermions.  
For $\Gamma < 1$, the role of the two operators is swapped and the situation 
is reversed: this symmetry is due to the self-duality of the model. 
These conclusions are confirmed by the explicit expressions 
of the operators in terms of the fermion fields given, for $\Gamma > 1$, 
by the normal ordered expressions~\cite{Jimbo}  
\begin{eqnarray}
&& \epsilon(r,t) = i \, :\psi_+(r,t) \, \psi_-(r,t):  \nonumber \\
&& \sigma(r,t) = \, : \psi_0(r,t) \, e^{\rho(r,t)} : 
\label{operatorsintermsofferions}
\\
&& \mu(r,t) = \, : e^{\rho(r,t)}: \nonumber \;,
\end{eqnarray} 
where $\rho(r,t)$ is a quadratic form of fermions  
\begin{eqnarray}
\rho(r,t) & = &  -i \int \frac{du}{2\pi u} \,\frac{du'}{2\pi u'} 
\frac{(u-u')}{u+u'+i \delta} \,\nonumber \\
& & \hspace{3mm} \times \, \hat\psi(u) \hat\psi(u') 
e^{-i(E+E') t + i (p+p') r} \;.
\end{eqnarray}
Notice that $\rho(r,t)$ is expressed by a non-local quantity of the fermion modes
which plays the role of the fermion string that accompanies the definition 
of the lattice operators (see Eq.~\eqref{stringfermion}). 

If we now set $u = e^{\beta}$, where $\beta$ is the rapidity, using the 
operatorial expressions above, one can easily check that, for $\Gamma > 1$, 
the matrix elements of the various operators (between the vacuum state $\ket{0}$ 
and the asymptotic states of $n$ fermions identified 
by their rapidities $\ket{\beta_1,\beta_2 ,\ldots,\beta_n}$) 
are given by~\cite{mussardobook,Alyosha}   
\[
\bra{0} \epsilon(0,0) \ket{\beta_1,\ldots,\beta_n} = 
\left\{  \begin{array}{cl}
i \Delta \, \sinh\frac{\beta_1 - \beta_2}{2} & {\rm for} \; n = 2 \\ 0 & {\rm otherwise} 
\end{array} \right.
\]
\[
\bra{0} \sigma(0,0) \ket{\beta_1,\ldots ,\beta_{2n+1}} = (i)^{2n+1}
\prod_{i < j}^{2n +1} \tanh\frac{\beta_i-\beta_j}{2} 
\]
\[ 
\bra{0} \mu(0,0) \ket{\beta_1,\ldots ,\beta_{2n}} = (i)^{2n}
\prod_{i < j}^{2n } \tanh\frac{\beta_i-\beta_j}{2} \;.
\]
These expressions of the matrix elements clearly show  that the order/disorder 
operators are non-local fields with respect to the fermion excitations: 
they couple to an arbitrary odd/even number of quasiparticles and they 
may be consequently regarded as strongly-interacting fields, even though the 
Hamiltonian~\eqref{DiracHamiltonian} is quadratic. 
On the other hand, the energy field associated to $\sigma^z$ is a local 
operator with respect to the fermions and it couples only to two-particle states. 

Another local field with respect to the fermion quasi-particle exitations 
is the density-of-kinks operator, whose lattice definition is  
 \beq
{\cal N} \equiv \frac{1}{L} \sum_j \frac{1 - \sigma^x_j \sigma^x_{j+1}}{2} \,.
\eeq
This operator is in fact at most quadratic in the fermion fields: this can be easily seen 
firstly using the Wigner-Jordan fermions to express 
$\sigma^x_j \sigma^x_{j+1} = \left[c^\dagger_j \, c_{j+1} + c^\dagger_j c^\dagger_{j+1} + h.c. \right]$, 
and secondly rewriting this expression in terms of the fermionic variables ${\cal A}_k$, by means
of the Bogoliubov transformation. Therefore this operator only couples to two-particle states.  

\subsection{Boundary state}

Before proceeding with the description of the dynamics following a quantum quench, 
it is worth discussing the representation of the initial state $\ket{\cal B}$ in terms of the quasiparticles
diagonalizing the final Hamiltonian and its connection to boundary states in statistical field
theory~\cite{ghoshalzam}. Despite the fact that this connection can be elucidated directly 
within the lattice model~\cite{silva08}, here we follow an elegant route employing the 
continuum formulation and the equations of motion for the Majorana fields:
\beq
\begin{array}{lcc}
\left(\partial_t + \partial_r\right) \psi_+(r,t) & = & \phantom{-} \Delta \, \psi_-(r,t) \,\, \phantom{.}\\
\left(\partial_t - \partial_r \right) \psi_-(r,t) & = & - \Delta \, \psi_+(r,t) \;.
\end{array}
\label{Diraceqs}
\eeq
In this formulation the quench process consists of an abrupt change 
of the mass, $\Delta_0 \rightarrow \Delta$, of the fermion field at $t=0$. 
Since the equations of motion~\eqref{Diraceqs} are of the first order in $\partial_t$, 
the fermion field should have no discontinuity at $t=0$ and therefore 
should satisfy the boundary condition 
\beq 
\psi^0_{\pm}(r,t=0) = \psi_{\pm}(r,t=0) \;,
\label{continuity}
\eeq
where $\psi^0_{\pm}(r,t)$ and $\psi_{\pm}(r,t)$ denote the fields relative
to the masses $\Delta_0$ and $\Delta$, respectively. 
Eq.~\eqref{continuity} implies a linear relation between the modes 
of the field before and after $t=0$. In order to compare directly 
with the lattice results, it is convenient to discretize the space 
variable in units of the lattice spacing $a$. 
In the following, focusing for simplicity on the case in which both 
$\Gamma_0,\Gamma>1$, let us write the mode 
expansion~\eqref{modeexpansionu} of the two fermionic fields as 
\beq
\psi_{\pm}(r,t)= \int_{BZ} \frac{dk}{2\pi}
\left[\alpha_{\pm}(k) {\cal A}(k) e^{-i E t + i k r} + h.c. \right] 
\label{modep}
\eeq
where the integral is over the first Brillouin zone $|k| < \pi/a$, 
$E = \sqrt{\Delta^2 + \tilde p^2}$ with $\tilde p = (2/a) \sin{(ka/2)}$, and 
$\alpha_{\pm}(k)=\omega^{\pm 1} \,\sqrt{\frac{E \pm \tilde{p}}{ 2 E}}$. 
Analogous expressions hold for the two components  
$\psi^0_{\pm}(r,t)$ expressed in terms of the modes ${\cal A}_0(k)$ and 
${\cal A}_0^{\dagger}(k)$, with $E_0 = \sqrt{\Delta_0^2 + \tilde p^2}$. 

Let us denote by $\ket{\cal B}$ the ground state of the fermion fields
$\psi^0_{\pm}(r,t)$ and by $\ket{0}$ the ground state of the field $\psi_{\pm}(r,t)$:  
these states are annihilated  by ${\cal A}_0(k)$ and ${\cal A}(k)$ respectively.
Extracting at $t=0$ the Fourier mode $\tilde\psi_{\pm}(k)$ of the fermion fields, 
defined by $\psi_{\pm}(r,0) = \int_{BZ} dk/(2\pi) \, \tilde\psi_{\pm}(k) e^{i k r}$,    
\beq
\begin{array}{ccc}
\tilde\psi_+(k) & = & \alpha_+(k) {\cal A}(k) + \bar\alpha_+(-k) {\cal A}^{\dagger}(-k) \\
\tilde\psi_-(k) & = & \alpha_-(k) {\cal A}(k) + \bar\alpha_-(-k) {\cal A}^{\dagger}(-k) 
\end{array}
\eeq
(with analogous expressions for $\psi_{\pm}^0$) and imposing the boundary
condition~\eqref{continuity}, one obtains the linear relations among the modes: 
\beq
\begin{array}{ccc}
{\cal A}(k) & = & {\cal U}(k) {\cal A}_0(k) - i {\cal V}(k) {\cal A}_0^{\dagger}(-k) \,\,\,, \\
{\cal A}^{\dagger}(k) & = & {\cal U}(k) {\cal A}_0^{\dagger}(k) + i {\cal V}(k) {\cal A}_0(-k)\,\,\,, 
\end{array}
\label{bogquench}
\eeq
where 
\begin{eqnarray*}
{\cal U}(k) & = & \frac{1}{2 \sqrt{E E_0}} \left[\sqrt{(E_0 + \tilde p)(E + \tilde p)}
 + \sqrt{(E_0 - \tilde p) (E - \tilde p)}\right] \\
{\cal V}(k) & = & \frac{1}{2 \sqrt{ E E_0}} \left[\sqrt{(E_0 - \tilde p)(E + \tilde p)}
 - \sqrt{(E_0 + \tilde p) (E - \tilde p)}\right]  \;.
\end{eqnarray*}
Notice that ${\cal U}(k) = {\cal U}(-k)$, ${\cal V}(k) = - {\cal V}(-k)$.
The inverse relations of Eqs.~\eqref{bogquench} are given by  
\beq
\begin{array}{ccc}
{\cal A}_0(k) & = & {\cal U}(k) {\cal A}(k) + i {\cal V}(k) {\cal A}^{\dagger}(-k) \;, \\
{\cal A}_0^{\dagger}(k) & = & {\cal U}(k) {\cal A}^{\dagger}(k) - i {\cal V}(k) {\cal A}(-k) \;. 
\end{array}
\label{bogquenchinv}
\eeq
Looking at these equations, we see that a quench in the mass of the 
fermion is simply equivalent to a Bogoliubov transformation of its 
modes -- transformation that is ruled by the functions ${\cal U}(k)$ and ${\cal V}(k)$.
The role of the initial state of the quench process is played by 
the ground state $\ket{\cal B}$ of the fermion fields $\psi^0_{\pm}(r,t)$. 

It is simple to work out the properties of the state $\ket{\cal B}$, hereafter called, 
for obvious reasons, {\em boundary state}~\cite{calabrese06,ghoshalzam}. 
Since this is the vacuum of the fermion $\psi^0(r,t)$, we can use the formula 
\beq 
\bra{\cal B} {\cal A}_0(k) {\cal A}_0^{\dagger}(k') \ket{\cal B} \,=\, 2 \pi  \delta(k-k') 
\eeq
and the Bogoliubov transformations~\eqref{bogquench} to easily compute 
the expectation values of modes ${\cal A}(k)$ and ${\cal A}^{\dagger}(k)$ 
on this state:
\begin{eqnarray}
 \langle {\cal B} | {\cal A}(k') {\cal A}(k) | {\cal B} \rangle  & = & 
2 \pi i \,{\cal U}(k') {\cal V}(k') \,\delta(k'+k) \nonumber \\
 \langle {\cal B} | {\cal A}(k') {\cal A}^{\dagger}(k) | {\cal B} \rangle  & = & 
2 \pi  \,{\cal U}^2(k') \, \delta(k'-k) 
\label{newvev}\\
 \langle {\cal B} | {\cal A}^{\dagger}(k') {\cal A}(k) | {\cal B} \rangle  & = &
2 \pi  \, {\cal V}^2(k')\, \delta(k'-k) \nonumber \\
\langle {\cal B} | {\cal A}^{\dagger}(k') {\cal A}^{\dagger}(k) | {\cal B} \rangle  & = & 
2 \pi i  \, {\cal U}(k') {\cal V}(k') \,\delta(k'+k) \nonumber \;.
\end{eqnarray}
%
\begin{figure}[!t]
  \begin{center}
    \includegraphics[scale=0.33]{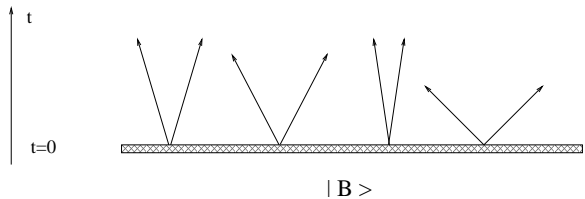}
    \vspace{0cm}
    \caption{With respect to the modes of the Hamiltonian at $t > 0$,
      the boundary state $\ket{\cal B}$ appears as a coherent superposition 
      of an infinite number of pairs of particles with equal and opposite momentum.}
    \label{fig:boundarystate}
  \end{center}
\end{figure}
%
However, we can gain new information on the nature of the boundary 
state $\ket{\cal B}$ if we directly express it in terms the oscillators 
${\cal A}(k)$ and ${\cal A}^{\dagger}(k)$. This can be done by using once again 
the condition that $\ket{\cal B}$ is annihilated by all the operators 
${\cal A}_0(p)$, but employing, this time, the first equation 
in~\eqref{bogquenchinv} for the operators ${\cal A}_0(p)$. 
In this way, we get the following infinite set of equations for the boundary state $\ket{\cal B}$:
\beq
\left[{\cal U}(k) {\cal A}(k) + i {\cal V}(k) {\cal A}^{\dagger}(-k) \right] \, \ket{\cal B} \,=\,0 \;, 
\label{eqsB}
\eeq
whose formal solution is a BCS-type state 
\beq
\ket{\cal B} \,=\,\prod_k \left[{\cal U}(k) + i {\cal V}(k) {\cal A}^{\dagger}(k) 
{\cal A}^{\dagger}(-k) \right] \ket{0} \;, 
\label{boundarystate}
\eeq
where $\ket{0}$ is the vacuum of the fermions $\psi_{\pm}(r,t)$.
Notice that, disregarding an overall normalization, the state $\ket{\cal B}$ 
can also be expressed as 
\beq
\ket{\cal B} = \exp\left[\frac{i}{2} \int_{BZ} \frac{dk}{2\pi} {\cal K}(k) {\cal A}^{\dagger}(k) 
{\cal A}^{\dagger}(-k) \right] \ket{0} \;,
\label{boundaryintegral}
\eeq
where ${\cal K}(k) = {\cal U}(k)/{\cal V}(k) = -{\cal K}(-k)$.
Therefore this belongs to the special class of initial states, recently 
studied in Ref.~[\onlinecite{fiorettomussardo}], represented by
a coherent superposition of particle pairs for which the long-time limit
of {\em one-point} functions is described by a generalized Gibbs ensemble. 
The expression~\eqref{boundarystate} or, equivalently~\eqref{boundaryintegral},
shows that, as a state of the Hilbert space of the fermion $\psi(r,t)$, 
the boundary state $\ket{\cal B}$ is a coherent superposition of pairs 
of particles with equal and opposite momentum (see Fig.~\ref{fig:boundarystate}).   
Despite the strong similarity of Eq.~\eqref{boundaryintegral} with the boundary 
states studied by Ghoshal and Zamolodchikov~\cite{ghoshalzam} in the context of 
integrable scattering theories, however notice that integrability in scattering 
theories imposes some extra conditions on the kernel ${\cal K}(k)$ which are not 
necessarily satisfied in the present case.

\section{Heisenberg evolution and two-point correlation functions} \label{sec:boundary}

Let us now turn to the dynamics. In the lattice formulation, 
this can be studied by looking at the Heisenberg evolution~\cite{mccoy1} 
of the fermionic operators $c_k(t)$. Setting 
\beq
\label{eq:ctime}
c_k (t) = u_k (t) {\cal A}^0_k - v^*_k (t) ({\cal A}^0_{-k})^\dagger \;,
\eeq
where ${\cal A}^0_k$ are the eigenmodes of the initial Hamiltonian $\Ham (\Gamma_0)$, 
and using the fact that the time evolution of the operators ${\cal A}_k$ 
diagonalizing the final Hamiltonian is trivial, it is easy to obtain that
\beq
\bar{c}_k(t)={\bf R}^{\dagger}_k(\Gamma)\;{\bf U}_k(t)\;{\bf R}_k(\Gamma)\;{\bf R}_k(\Gamma_0) \;\bar{\cal A}^0_k \;,
\eeq
where 
\beq
{\bf U}_k(t)=\left(
\begin{array}{cc} e^{-i\epsilon_k t} & 0 \\ 0 & e^{i\epsilon_k t} \end{array} \right) \;.
\eeq
The relevant dynamical observables can then be evaluated on the time
evolved state, Eq.~\eqref{eq:psit}, by using the Heisenberg picture and 
Eq.~\eqref{eq:ctime}, after writing them in terms of Jordan-Wigner fermions.
We are now ready to provide some technical 
details on the computation of the two-point spin correlation functions.

\subsection{Order-parameter spin correlation functions} \label{subsec:XXtechnics}

The time-dependent correlation function of the order parameter is defined as:
\beq
   \rho^{xx} (t,r) \equiv \bra{\cal B} \sigma^x_{m+r} (t_0+t) \, \sigma^x_m (t_0) \ket{\cal B} \;,
   \label{eq:XXcorr}
\eeq
where $t_0$ denotes the waiting time after the quench.
Due to the translational invariance of the system, Eq.~\eqref{eq:XXcorr}
does not depend on $m$, and we can choose $m=1$ without loss of generality.
The operator $\sigma^x_{m+r} (t_0+t) \, \sigma^x_m (t_0)$ connects states with different
$c$-fermionic parity; therefore it cannot be simply evaluated using
Jordan-Wigner fermions in the even Hamiltonian sector $\Ham^+$.
This problem can be circumvented~\cite{mccoy4} by considering a four-spin
correlation function on a chain of length $L$, with $r<L/2$:
\beq
\begin{array}{rl}
C^{xx} (t,r;L) = & 
\bra{\cal B} \sigma^x_{L-r+1} (t_0+t) \, \sigma^x_{1} (t_0) \, \vspace*{1mm} \\
 & \hspace*{3.mm} \times \sigma^x_{\frac{L}{2}+1} (t_0+t) \, \sigma^x_{\frac{L}{2}-r+1} (t_0) \ket{\cal B} \;.
\end{array}
\label{eq:fourpoint}
\eeq
By using the cluster property and taking the thermodynamic limit~\cite{mccoy4},
one can show that this function reduces to the square of the correlation
function $\rho^{xx} (t)$:
\beq
\big( \rho^{xx} (t,r) \big)^2 = \lim_{L \to \infty} C^{xx} (t,r;L) \;;
\eeq
the crucial advantage of this strategy is that the four-point
correlator in Eq.~\eqref{eq:fourpoint} conserves the $c$-fermion parity, and can
therefore be evaluated in the (antiperiodic) even fermionic sector~\cite{mccoy4}.

Following Refs.~[\onlinecite{mccoy4,lieb61}], Eq.~\eqref{eq:fourpoint} 
can be written as a Pfaffian.
We first write it in terms of Jordan-Wigner fermions, by simply using
$\sigma^x_j = e^{i \pi \sum_{k<j} n_k} (c^\dagger_j + c_j)$;
then, after defining $A_j (t) \equiv c_j^\dagger (t) + c_j (t)$ and
$B_j (t) \equiv c_j^\dagger (t) - c_j (t)$, we get:
\beq
\begin{array}{l}
C^{xx} (t,r;L) = \bra{\cal B} 
[ B_{\frac{L}{2}+1} (t_0+t)  \cdots  B_{L-r} (t_0+t)                ] \vspace{1mm} \\
\hspace*{23.7mm} \times [ A_{\frac{L}{2}+2} (t_0+t)  \cdots  A_{L-r+1} (t_0+t)              ] \vspace{1mm} \\
\hspace*{4.mm} \times
[ B_1 (t_0)                \cdots  B_{\frac{L}{2}-r} (t_0)      ] \times
[ A_2 (t_0)                \cdots  A_{\frac{L}{2}-r+1} (t_0)    ]   \ket{\cal B} \,,
\end{array}
\label{eq:ABcorr}
\eeq
in which we also used the equalities $(1- 2 n_j) = A_j B_j =-B_jA_j$ 
and $\{ A_j , B_l \} =0 \;\; \forall j,l$ (here $\{ \cdot , \cdot \}$ denotes the anticommutator).
An application of Wick's theorem then leads to the following expression,
where the square of the correlator $C^{xx}$ is written in terms of 
the determinant of a $(2L -4r) \times (2L - 4r)$ matrix:
\begin{widetext}
\beq
\begin{array}{ll}
\big( C^{xx} (t,r;L) \big)^2 = \vspace*{2mm} \\ 
\hspace*{4mm} \left| \begin{array}{cccc}
\left<  B_{[j_1]} (t_0+t) B_{[l_1]} (t_0+t) \right> &  \left< B_{[j_1]} (t_0+t) A_{[l_2]} (t_0+t) \right> &
\left<  B_{[j_1]} (t_0+t) B_{[l_3]} (t_0) \right> &  \left< B_{[j_1]} (t_0+t) A_{[l_4]} (t_0) \right> \vspace{1.mm}\\
-\left< B_{[l_1]} (t_0+t) A_{[j_2]} (t_0+t) \right> &  \left< A_{[j_2]} (t_0+t) A_{[l_2]} (t_0+t) \right> &
\left<  A_{[j_2]} (t_0+t) B_{[l_3]} (t_0) \right> &  \left< A_{[j_2]} (t_0+t) A_{[l_4]} (t_0) \right> \vspace{1.mm}\\
-\left< B_{[l_1]} (t_0+t) B_{[j_3]} (t_0) \right> & -\left< A_{[l_2]} (t_0+t) B_{[j_3]} (t_0) \right> &
\left<  B_{[j_3]} (t_0) B_{[l_3]} (t_0) \right> &  \left< B_{[j_3]} (t_0) A_{[l_4]} (t_0) \right> \vspace{1.mm}\\
-\left< B_{[l_1]} (t_0+t) A_{[j_4]} (t_0) \right> & -\left< A_{[l_2]} (t_0+t) A_{[j_4]} (t_0) \right> &
-\left< B_{[l_3]} (t_0) A_{[j_4]} (t_0) \right> &  \left< A_{[j_4]} (t_0) A_{[l_4]} (t_0) \right>
\end{array} \right|
\end{array}
\label{eq:toeplitz}
\eeq
\beq
{\rm with} \quad \left\{ \begin{array}{l}
\left< A_j (t_1) A_l (t_2) \right> = \displaystyle \frac{1}{L} \sum_k e^{i k (j-l)}
\big( u_k (t_1) + v_k (t_1) \big) \big( u_k^* (t_2) + v_k^* (t_2) \big) \\
\left< A_j (t_1) B_l (t_2) \right> = \displaystyle \frac{1}{L} \sum_k e^{i k (j-l)}
\big( u_k (t_1) + v_k (t_1) \big) \big( u_k^* (t_2) - v_k^* (t_2) \big) \\
\left< B_j (t_1) A_l (t_2) \right> = \displaystyle \frac{1}{L} \sum_k e^{i k (j-l)}
\big( v_k (t_1) - u_k (t_1) \big) \big( u_k^* (t_2) + v_k^* (t_2) \big) \\
\left< B_j (t_1) B_l (t_2) \right> = \displaystyle \frac{1}{L} \sum_k e^{i k (j-l)}
\big( u_k (t_1) - v_k (t_1) \big) \big( v_k^* (t_2) - u_k^* (t_2) \big) \;.
\end{array} \right.
\label{eq:aacomput}
\eeq
\end{widetext}
In Eq.~\eqref{eq:toeplitz} the two indexes $j,l$ have subscripts which
indicate their range: an index with label $_1$ runs from $\frac{L}{2}+1$ to $L-r$,
with $_2$ runs from $\frac{L}{2}+2$ to $L-r+1$,
with $_3$ runs from $1$ to $\frac{L}{2}-r$,
with $_4$ runs from $2$ to $\frac{L}{2}-r+1$.
Each of the 16 submatrices in~\eqref{eq:toeplitz} is therefore of dimension
$(\frac{L}{2}-r) \times (\frac{L}{2}-r)$, and can be explicitly evaluated by
substituting the proper definitions of the $A_j$'s and $B_j$'s and then working
in momentum space, as detailed in Eqs.~\eqref{eq:aacomput}
[$k$-sums are taken for fermion antiperiodic boundary conditions, Eq.~\eqref{eq:abc}].
All the diagonal entries of the matrix in~\eqref{eq:toeplitz} are zero,
since they do not enter the contractions in Wick's 
theorem~\footnote{
We point out that in a recent paper 
[J.~H.~H.~Perk and H.~Au-Yang, J. Stat. Phys. {\bf 135}, 599 (2009)]
an alternative method for evaluating such two-point correlations of the order parameter,
based on the resolution of a pair of Toda-type nonlinear differential equations,
has been put forward.}.

\subsection{Transverse-field spin correlation functions}  \label{subsec:ZZtechnics}

We now briefly sketch how to compute the two-point correlation functions of 
the transverse-field magnetization:
\beq \label{eq:ZZcorr}
\rho^{zz} (t,r) \equiv \bra{\cal B} \sigma^z_{m+r} (t_0+t) \, \sigma^z_m (t_0) \ket{\cal B} \;.
\eeq
As before, since we are using periodic boundary conditions, we can take $m=1$.
Contrary to the order parameter correlation function, the operator
$\sigma^z_{m+r} (t_0+t) \, \sigma^z_m (t_0)$ conserves the $c$-fermion parity,
therefore one can rewrite it using Jordan-Wigner fermions
in the even Hamiltonian sector $\Ham^+$, and the computation becomes simple.
Using $\sigma^z_j = 2 \, c^\dagger_j c_j - 1$, one gets
$\rho^{zz} (t,r) = \bra{\cal B} ( 2 c^\dagger_{r+1} (t_0+t) \, c_{r+1} (t_0+t) - 1 ) \times
( 2 c^\dagger_1 (t_0) \, c_1 (t_0) - 1 ) \ket{\cal B}$.
Switching to momentum space, and using Eq.~\eqref{eq:ctime}
together with the fact that $\ket{\cal B}$ is the vacuum for 
${\cal A}^0_k$ particles, the following expression can be easily obtained:
\beq
\label{eq:ZZcorruv}
\begin{array}{l}
\displaystyle \rho^{zz} (t,r) = 1- \frac{2}{L} \sum_k
\Big[ \vert v_k (t_0) \vert^2 + \vert v_k (t_0+t) \vert^2 \Big] \\
\displaystyle \hspace*{8.mm} + \frac{4}{L^2} \sum_{k,l} \Big[
e^{ir(l-k)} \big( v_k(t_0+t) \, v_k^*(t_0) \, u_l(t_0+t) \, u_l^*(t_0) \\
\displaystyle \hspace*{23.mm} + v_k(t_0+t) \, v_l^*(t_0) \, u_l(t_0+t) \, u_k^*(t_0) \big) \vspace*{0.5mm}  \\
\displaystyle \hspace*{23.mm} + \vert v_k (t_0+t) \vert^2 \: \vert v_l (t_0) \vert^2  \Big] \;.
\end{array}
\eeq

\subsection{Density of kinks}  \label{subsec:KINKtechnics}

The density of kinks for the Ising model~\eqref{eq:model} is defined
by the operator 
\beq
{\cal N} \equiv \frac{1}{L} \sum_j \frac{1 - \sigma^x_j \sigma^x_{j+1}}{2} \,.
\eeq
Its expectation value $\rho^{\cal N} (t)$ at a certain time $t$ on the 
boundary state $\ket{\cal B}$ is strictly related to the nearest-neighbor
order-parameter correlation function. For periodic boundary conditions
this can be simply written as: 
\beq
\rho^{\cal N} (t) =\frac{1 - \bra{\cal B} \sigma^x_m(t) \, \sigma^x_{m+1}(t) \ket{\cal B}}{2}
= \frac{1 - \rho^{xx} (0,1)}{2} \, ,
\eeq
where $\rho^{xx} (0,1)$ is evaluated after a waiting time $t_0 \equiv t$.
Proceeding in an analogous way as in Sec.~\ref{subsec:XXtechnics}-\ref{subsec:ZZtechnics},
this can be explicitly rewritten as:
\barr
\rho^{\cal N} (t) &=& \frac{1}{2L} \sum_k
\Big[ 1 - \big( |v_k(t)|^2 - |u_k(t)|^2 \big) \cos k \nonumber\\
 && \hspace*{0.5cm}
- i \big( u_k(t)v_k^{\ast}(t) - u_k^{\ast}(t)v_k(t) \big) \sin k
\Big] \;.
\label{eq:KINKcomput}
\earr

Notice that Eqs.~\eqref{eq:aacomput},~\eqref{eq:ZZcorruv} and~\eqref{eq:KINKcomput} 
refer to a finite system of $L$ spins; the thermodynamic limit can be formally 
attained by taking the continuum limit and
replacing the sums with integrals: $1/L\sum_k \to \int dk/(2\pi)$.

\section{Thermal relaxation of non-local operators: order parameter correlations}
\label{sec:XXquench}

In this section we discuss in details the behavior of the correlation function of the order parameter
$\rho^{xx}_Q (t,r)$ for a quantum quench of the transverse field from $\Gamma_0$ to $\Gamma$,
focusing on the asymptotic relaxation for long times ($t \to \infty$) and for long distances ($r \to \infty$).  
The correlation function in Eq.~\eqref{eq:XXcorr} can be explicitly written as
\beq
\rho^{xx}_Q (t,r) = \bra{{\cal B}} e^{i \Ham(\Gamma) (t_0+t)} \sigma^x_{m+r}
e^{-i \Ham(\Gamma) t} \sigma^x_m e^{-i \Ham(\Gamma) t_0}  \ket{\cal B} \;. 
\label{eq:XXquench}
\eeq
This correlator can be calculated following the prescriptions given in Sec.~\ref{subsec:XXtechnics},
i.e., by numerically computing the Toeplitz determinant in Eq.~\eqref{eq:toeplitz}.

\subsection{Time-dependent correlations} \label{subsec:XXtimedep}

Let us start our analysis with the autocorrelation functions ($r=0$).
For the system at equilibrium, time translation invariance implies that 
the autocorrelation functions do not depend on the waiting time $t_0$ in 
Eq.~\eqref{eq:XXquench}. Quite remarkably, we found that, as long as one considers
$\rho^{xx}_Q(t,0)$ at large times, this is true also for the nonequilibrium case.
Therefore, in this subsection we set for simplicity $t_0=0$.
The relevant features emerging for the case $r=0$, $t_0=0$ have been already 
elucidated by some of us in Ref.~[\onlinecite{paper1}].
Here we summarize and discuss in more details the results anticipated in that paper.

\begin{figure}[!t]
  \begin{center}
    \includegraphics[scale=0.3]{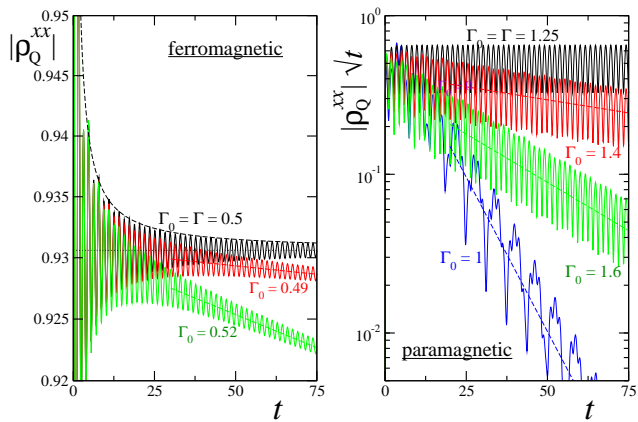}
    \caption{(color online). Absolute value of the on-site time dependent 
      correlation function $\rho^{xx}_Q (t,0)$ for quenches terminating 
      in the ferromagnetic ($\Gamma = 0.5$ --left panel),
      and in the paramagnetic phase ($\Gamma = 1.25$ --right panel).
      The various curves are obtained for different
      values of $\Gamma_0$, as indicated in the graphs.
      The black curve corresponds to the equilibrium case
      (in the left panel, the horizontal dotted line indicates 
      the asymptotic value $\langle \sigma^x \rangle^2$,
      while the dashed one denotes the power-law envelope $\sim t^{-1}$).
      Colored curves stand for different quenches $\Gamma_0 \neq \Gamma$;
      the straight dashed lines indicate the respective leading exponential decays.
      In the paramagnetic case, curves are rescaled by the 
      zero-temperature $1/\sqrt{t}$ prefactor, analogous to the
      $K(t,0)$ term in Eq.~\eqref{eq:xxT_para}.}
    \label{fig:CorrXX}
  \end{center}
\end{figure}

The envelope of the autocorrelation function
$\rho_Q^{xx} (t,0) \equiv \langle {\cal B} |\sigma^x_m(t) \sigma^x_m(0) | {\cal B} \rangle$
relaxes exponentially to zero in time:
\beq
\rho_Q^{xx} (t,0) \simeq e^{-t/\tau^\varphi_Q} \;.
\eeq
This applies both to quenches ($\Gamma_0 \neq \Gamma$) ending in the ferromagnetic phase ($\Gamma<1$) or, 
as already pointed out by Calabrese and Cardy~\cite{calabrese06}, at criticality ($\Gamma=1$). 
As it can be clearly seen from the left panel of Fig.~\ref{fig:CorrXX}, as soon as $\Gamma \neq \Gamma_0$,
a sharp contrast with the zero-temperature behavior in equilibrium becomes evident (black curve).
Indeed, at zero temperature and at equilibrium, the correlator 
$\rho^{xx}_{T=0}(t,r)$ asymptotically tends, both in time and in space,
to the square of the order parameter $\langle \sigma^x \rangle$~\cite{pfeuty70}.
In that case, the asymptotic value $\langle \sigma^x \rangle^2$ is not reached exponentially, 
but through an oscillatory power-law decay with an exponent $\alpha$ 
depending on the phase of the system ($\alpha=1$ in the ferromagnetic phase, 
$\alpha = 1/4$ at criticality, and $\alpha = 1/2$ in the paramagnetic phase~\cite{mccoy4}).

Drawing a parallel with the equilibrium physics, the exponential decay
to zero observed for quantum quenches is reminiscent of the exponential 
decay of the auto-correlation function observed in equilibrium at 
finite temperatures $T>0$. The rate $\tau^\varphi_Q$ can be 
identified with the non-equilibrium analogue of the phase coherence time. 
In the equilibrium case, a semiclassical analysis due to 
Sachdev and Young~\cite{sachdev97} shows that, for $T \ll \Delta$ 
and in the ferromagnetic phase, one has:
\beq \label{eq:xxT_ferro}
   \rho^{xx}_T (t,r) \simeq N_0^2 \, R_T(t,r) \;,
\eeq
where $N_0$ is the vacuum expectation value of the order parameter 
$N_0= \langle \sigma^x \rangle$ and $R_T(t,r)$ is a relaxation function 
describing thermal excitations:
\beq \label{eq:RelaxT}
   R_T(t,r) = \exp \left( - \int \frac{dk}{\pi} e^{-\epsilon_k/T} \vert r - v_k t \vert \right) \;.
\eeq
Here $v_k = \partial_k \epsilon_k$ is the velocity of the thermally excited
quasiparticles (kinks) and $e^{-\epsilon_k/T}$ their Boltzmann weight.
From Eq.~\eqref{eq:RelaxT}, one can readily extract the basic time- and length-scales of the system
in the ferromagnetic region, the phase coherence time $\tau^\varphi_T$ 
and the correlation length $\xi^\varphi_T$. Indeed, $R_T(t,0) = e^{-t/\tau^\varphi_T}$ with:
\beq \label{eq:coherenceT} 
\tau^\varphi_T = \left( \int \! \frac{dk}{\pi} e^{-\epsilon_k/T} |v_k| \right)^{-1} 
\approx \frac{\pi}{2 T} e^{\Delta/T} \;,
\eeq
and $R_T(0,r) = e^{-r/\xi^\varphi_T}$ with 
\beq \label{eq:coherenceR}
\xi^\varphi_T = c \left( \int \! \frac{dk}{\pi} e^{-\epsilon_k/T} \right)^{-1} 
\approx \sqrt{\frac{\pi c}{2T \Delta}} e^{\Delta/T} \;,
\eeq
where $c$ is the ``speed of light'', given in terms of the bandwidth $J$ and of the lattice constant
$a$ by $c=2\;J\;a\;\sqrt{\Gamma}$.
Notice that in the critical region ($T \sim \Delta$) quasiparticles are not well defined and
semiclassical arguments are not applicable. 
However, on the basis of an analysis of the continuum scaling limit one obtains
\beq
\tau^\varphi_T \;,\; \xi^\varphi_T \sim \frac{1}{T} \;.
\label{eq:xxT_crit}
\eeq
This behavior is confirmed by an exact analytic calculation on the lattice~\cite{deift94},
giving $\tau^\varphi_T = \frac{8}{\pi T}$.

The parallel between equilibrium behavior of the autocorrelation function at finite $T$,
and the behavior after a quench is even more astounding when we consider 
quenches ending in the paramagnetic phase ($\Gamma > 1$), where the equilibrium analysis
(in spite of the fact that quasiparticles are no longer kinks) predicts a
very similar structure~\cite{sachdev97}: 
\beq \label{eq:xxT_para}
   \rho^{xx}_T (t,r) \simeq K(t,r) \cdot R_T(t,r) \;,
\eeq
with $R_T(t,r)$ still given by Eq.~\eqref{eq:RelaxT}, while $K(r,t)$ 
is an oscillatory decaying function determined by the quantum fluctuations 
in the ground state. 
For the Ising chain, $K(t,r) \sim K_0 (\Delta \sqrt{r^2 - \Gamma t^2}/c)$,
where $K_0$ is the modified Bessel function, which oscillates and decays 
as $1/\sqrt{t}$ for large $t$, while it decays exponentially as $e^{-r \Delta/c}$ for
large $r$. 
The data for the autocorrelator $\rho^{xx}_Q(t,0)$ as a function of
time with final $\Gamma>1$ show pronounced oscillations as in the equilibrium case, and
the exponential drop clearly emerges only if $\rho^{xx}_Q(t,0)$
is multiplied by $\sqrt{t}$, the leading decay
of $K(t,0)$ (see the right panel of Fig.~\ref{fig:CorrXX}).

Given a generic quench from $\Gamma_0$ to $\Gamma$, one can
numerically extract a relaxation rate $\tau^\varphi_Q$.
As shown in Fig.2 of Ref.~[\onlinecite{paper1}], $\tau^\varphi_Q$ 
decreases as the quench ``strength'' $\vert \Gamma_0 - \Gamma \vert$ increases, 
implying that the more the system
is driven out of equilibrium, the faster correlations decay in time.
In the limiting case of equilibrium ($\Gamma\to \Gamma_0$), 
$\tau^\varphi_Q \to \infty$ and the exponential relaxation turns into a power-law.
Notice that, even though the decay rate of correlations depends on 
the strength of the quench, the value reached asymptotically is always zero, 
$\rho^{xx}_Q (t,0) \stackrel{t \to \infty}{\longrightarrow} 0$,
irrespective of the values of $\Gamma_0$ and $\Gamma$.

We can understand these results as follows: the effect of a quench 
from $\Gamma_0$ to $\Gamma$ consists of injecting in the system an
extensive amount of energy. As seen in Sec~\ref{sec:boundary}, this leads 
to the generation of a finite density of quasi-particle excitations ${\cal A}_k$, 
with a non-equilibrium occupation 
\beq \label{f_k:eqn}
f_k = \bra{\cal B} {\cal A}^\dagger_k {\cal A}_k \ket{\cal B} \;.
\eeq
The farther is $\Gamma_0$ from $\Gamma$, and therefore the greater is the injected energy, 
the higher is the density of quasiparticles created.
This is qualitatively very similar to the physics behind the decay of the autocorrelation function 
at equilibrium, if we consider the fact that the greater is the temperature $T$,
the higher is the density of quasiparticles in the system. 
It is therefore quite natural to describe the dependence of the relaxation rate on the 
initial state and on the final Hamiltonian in terms of an {\it effective temperature}~\cite{paper1}. 
The most natural way to define the effective temperature $T_{\rm eff}$ consists in 
comparing the energy of the initial state $\bra{\cal B} \Ham(\Gamma) \ket{\cal B}$
to that of a fictitious thermal state relative to the final Hamiltonian $\Ham(\Gamma)$
\beq
\langle \Ham(\Gamma) \rangle_{T_{\rm eff}} =
\sum_{k>0} \epsilon_k^\Gamma \Big( n_k (T_{\rm eff}) + n_{-k} (T_{\rm eff}) -1 \Big) \;,
\label{eq:encalc}
\eeq
where $n_k (T_{\rm eff}) = (1+e^{\epsilon_k^\Gamma/T_{\rm eff}})^{-1}$ is the Fermi 
distribution function for the quasiparticles ${\cal A}_k$ diagonalizing $\Ham(\Gamma)$
[see Eq.~\eqref{eq:hamdiag}].
The effective temperature is therefore determined by the implicit equation:
\beq \label{eq:Teff}
\bra{\cal B} \Ham(\Gamma) \ket{\cal B} = \langle \Ham(\Gamma) \rangle_{T_{\rm eff}} \;,
\eeq
or, in a completely equivalent way, by:
\begin{equation} \label{eq:Teff_bis}
\sum_{k>0} \epsilon_k^\Gamma f_k  = \sum_{k>0} \epsilon_k^\Gamma n_k (T_{\rm eff}) \;.
\end{equation}
For given $\Gamma_0$ and $\Gamma$ this equation always admits a single solution.

Quite remarkably, $\tau^\varphi_Q$ appears to be univocally determined
by $T_{\rm eff}$ and by the energy gap $\Delta(\Gamma)$ in the final state.
This means that two different quenches starting and ending at different
$\Gamma_0$ and $\Gamma$, but having the same effective temperature and the
same $\Delta(\Gamma)$ will show the same $\tau^{\varphi}_Q$, irrespective of whether the two final
$\Gamma$ are both in the same region of the phase diagram or, for example, one in the  
paramagnetic and the other in the ferromagnetic region.
Thus the fine details of the initial conditions are not important, while the only 
key physical parameter is the effective temperature of the initial state.

\begin{figure}[!t]
  \begin{center}
    \includegraphics[scale=0.34]{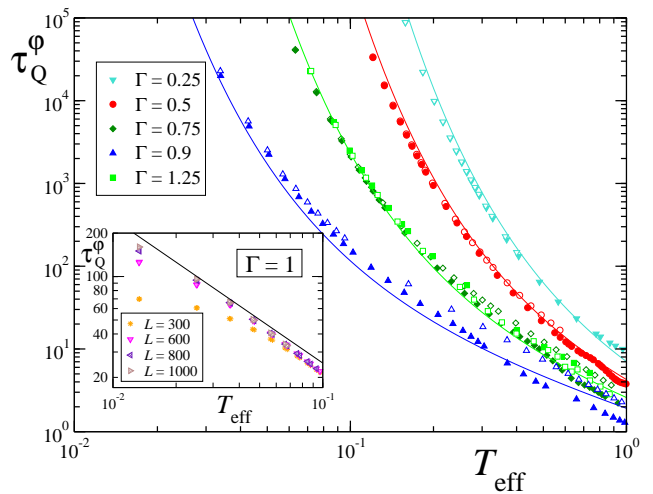}
    \caption{(color online). Phase coherence time as a function of
    the effective temperature, in the off-critical region for different
    values of $\Gamma$ (main panel) and at criticality (inset).
    Symbols refer to $\tau^\varphi_Q$ (here, as in Fig.~\ref{fig:Intro_thermal},
    empty symbols stand for quenches starting with $\Gamma_0 < \Gamma$; 
    filled ones are for $\Gamma_0 > \Gamma$),
    while continuous curves denote the equilibrium values
    at finite temperatures $\tau^\varphi_T$, Eq.~\eqref{eq:coherenceT}.
    For the system at criticality, we performed a finite-size scaling
    of $\tau^\varphi_Q$ to study the asymptotic agreement with the
    equilibrium law $\tau^\varphi_T = \frac{8}{\pi T}$ (black line).
    [data for $\Gamma = 0.5, \, 0.75, \, 1, \, 1.25$ acknowledged from
    Fig.~3 of Ref.~[\onlinecite{paper1}]].}
    \label{fig:Tau_Temp}
  \end{center}
\end{figure}

As one can see from the main panel of Fig.~\ref{fig:Tau_Temp},
away from criticality the phase coherence time $\tau^\varphi_Q$ (symbols) nicely follows, qualitatively
and even quantitatively, the curve describing the equilibrium phase coherence time 
$\tau^\varphi_{T=T_{\rm eff}}$ at the effective temperature $T_{\rm eff}$ [see Eq.~\eqref{eq:coherenceT}]. 
This statement is true also at the critical point, $\Gamma=1$, where, however,
a finite-size scaling is mandatory in order to capture the $T_{\rm eff}^{-1}$ behavior 
predicted at equilibrium. 
Indeed, while the comparison of $\tau^{\varphi}_Q$ with $\tau^\varphi_{T=T_{\rm eff}}$ systematically improves 
as $T_{\rm eff}$ is decreased, 
the long wavelength modes become increasingly important at low effective temperatures, thus making 
boundary effects more visible.
Numerical data are shown in the inset of Fig.~\ref{fig:Tau_Temp};
notably, after a finite-size scaling, the data not only follow the $T_{\rm eff}^{-1}$ law 
but also appear to have the same prefactor $8/\pi$ analytically predicted in Ref.~[\onlinecite{deift94}]
for a system at equilibrium.

Since the parallel between quench dynamics and thermal behavior works
better at low effective temperatures, one might be tempted to fix the effective
temperature directly in the long-wavelength scaling limit. However, particular care
should be taken in this case. Indeed, the energy of this state is easily computed 
using the matrix elements~\eqref{newvev} and is given by 
\beq
h_{\cal B} = \bra{\cal B} \Ham \ket{\cal B} \,=\, L \,\int_{BZ} \frac{dk}{2\pi} 
E(k) \, {\cal V}^2(k) \;,
\label{energyboundarystate}
\eeq
where the explicit length $L$ of the system here shows 
that we are in presence of a global quench. The integral above is finite 
as long as the lattice spacing $a$ is finite, whereas it is logarithmic 
divergent as $\log(1/a)$ when $a\rightarrow 0$. The simplest way to see 
the appearance of this logarithmic divergence when $a \rightarrow 0$ 
is to observe that, in the continuum limit, the mass couples to the operator 
$i :\psi_+(x) \psi_-(x):$ normal ordered with respect to the mass $m$ of the field. 
If we start with $m_0$ and change the mass $m_0 \rightarrow m=m_0 +\delta m$, 
this operator acquires a vacuum expectation value, and the energy density 
per unit length changes as  
\[
i \, \delta m \, \bra{\cal B} \psi_+(x) \psi_-(x) \ket{\cal B} \,=\,
(\delta m)^2 \,\int_0^{\infty} \frac{dp}{2\pi}\,\frac{p^2}{(p^2 + \Delta^2)^{3/2}} \;.
\]
This quantity, which is logarithmic divergent, matches with the limiting 
expression for $a \rightarrow 0$ of the integral~\eqref{energyboundarystate}, 
expanded at the same order in $\delta m$. This implies that, keeping fixed 
the energy density per unit length $h_{\cal B}/L$ of the boundary state while 
going to the continuum limit $a \rightarrow 0$, one is forced to consider 
only quenches where the differences in the masses are logarithmically close each other, 
$m-m_0 \sim 1/\log(1/a)$. With this adjustment of the masses, the physics 
of the quench process is invariant under a change of the lattice spacing.  

\subsection{Space-dependent correlations} \label{subsec:XXspacedep}

A similar analysis can be performed for the space dependence of the 
order parameter correlation functions~\eqref{eq:XXquench}, 
$\rho_Q^{xx} (0,r) \equiv \bra{\cal B} \sigma^x_{m+r}(t_0) 
\sigma^x_m(t_0) \ket{\cal B}$,
for sufficiently large times $t_0$ after the quantum quench.
As for the time autocorrelations, we find that this correlator displays an
exponential decay to zero as a function of the distance $r$:
\beq
\rho_Q^{xx} (0,r) \simeq {\tilde K}(r)\;e^{-r/\xi^\varphi_Q} \;.
\label{eq:rhoXXquench_para}
\eeq
This allows us to define the correlation length $\xi^\varphi_Q$ for a system after
the quench, which is in general a function of the quenching parameters $\Gamma_0$ and $\Gamma$. 

\begin{figure}[!t]
  \begin{center}
    \includegraphics[scale=0.32]{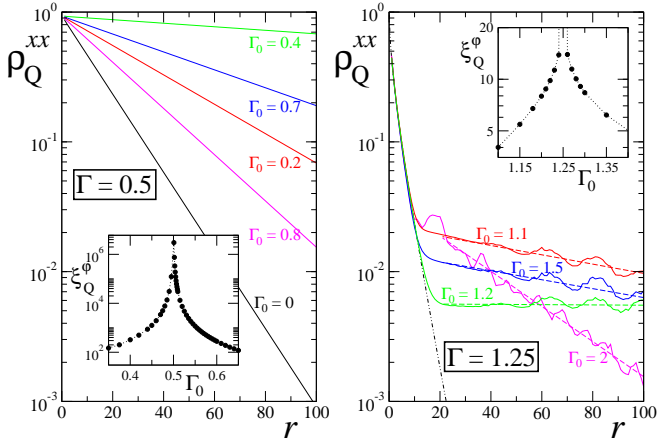}
    \caption{(color online). Space-dependent correlation function
      $\rho^{xx}_Q(0,r)$ for quenches in the ferromagnetic phase
      (left panel, $\Gamma = 0.5$) and in the paramagnetic phase
      (right panel, $\Gamma = 1.25$).
      The various data sets are for different values of $\Gamma_0$,
      as explicitly indicated near each curve.
      On the right panel, dashed colored lines are exponential fits
      of the curves in the large-$r$ limit, while the dotted-dashed black curve
      represents ${\cal K} (r)$ at the corresponding $\Gamma$.
      In the insets we plot the correlation length $\xi^\varphi_Q$
      as a function of $\Gamma_0$, as extracted from the exponential decay 
      in the space of correlations.}
    \label{fig:Ferro_xi_gamma}
  \end{center}
\end{figure}

The behavior of $\rho_Q^{xx} (0,r)$ as a function of the distance is explicitly
shown in Fig.~\ref{fig:Ferro_xi_gamma}.
For quenches in the ferromagnetic phase, one clearly identifies 
a pure exponential decay with $r$, as illustrated in the left panel of the figure;
therefore, the decay rate $\xi^\varphi_Q$ can be easily extracted 
by fitting the numerical data~\footnote{We checked that the exponential decay rates
do not depend on the waiting time $t_0$, provided this is sufficiently large.}.
This is analogous to the equilibrium case, Eq.~\eqref{eq:xxT_ferro},
as already observed for the time-dependent autocorrelation functions.
In the same way as for $\tau^\varphi_Q$, the correlation length
$\xi^\varphi_Q$ increases as the quench strength decreases and eventually diverges for $\Gamma=\Gamma_0$
(equilibrium, at $T=0$), where at large distances $\rho_{T=0}(0,r)$ tends to 
$\langle \sigma^x \rangle^2$ and not to zero;
the dependence of $\xi^\varphi_Q$ on $\Gamma_0$, for fixed $\Gamma$, is shown
in the inset of Fig.~\ref{fig:Ferro_xi_gamma}.

In the paramagnetic phase the situation is more subtle, as one can see
from the right panel of Fig.~\ref{fig:Ferro_xi_gamma}.
Here the behavior of $\rho_Q^{xx} (0,r)$ at short distances is 
dictated by the prefactor ${\tilde K}(r)$ which decays exponentially
to a constant ${\cal C} \neq 0$, ${\tilde K}(r) \approx {\cal C}+\exp[-\Delta r/c]$
(a similar behavior is present in equilibrium, where $K(0,r) \sim K_0 (\Delta r/c)$,
though with ${\cal C}=0$).
For distances $r \gg c/\Delta$, where ${\tilde K}\approx {\cal C}$,  the observed exponential 
decay univocally defines $\xi^{\varphi}_{Q}$. As a function of $T_{\rm eff}$ and $\Delta$, 
this appears to qualitatively follow the law in Eq.~\eqref{eq:coherenceR}, as in equilibrium.

\begin{figure}[!t]
  \begin{center}
    \includegraphics[scale=0.32]{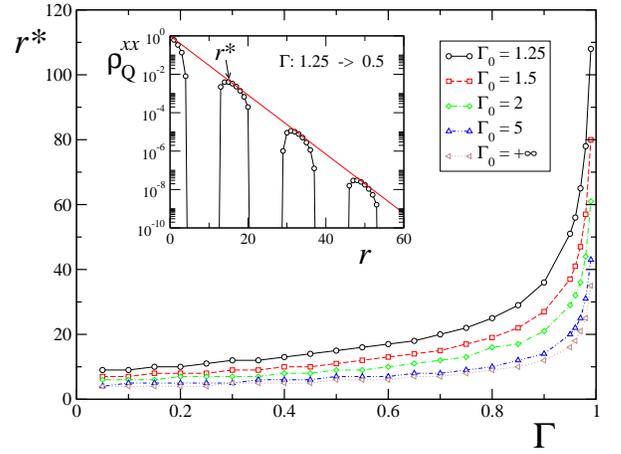}
    \caption{(color online) Distance $r^*$ at which the first maximum 
      in the oscillations for the spatial correlations crossing the
      critical point from the disordered phase appears, 
      as a function of $\Gamma<1$ and for fixed $\Gamma_0>1$. 
      Such value is a measure of the period of oscillations.
      In the inset we show $\rho^{xx}_Q(r)$ for a 
      paramagnetic-to-ferromagnetic quench.
      The oscillations are superimposed to an exponential decay
      behaving thermally,
      which is depicted with a straight red line.}
    \label{fig:recurr2}
  \end{center}
\end{figure}

\begin{figure}[!t]
  \begin{center}
    \includegraphics[scale=0.34]{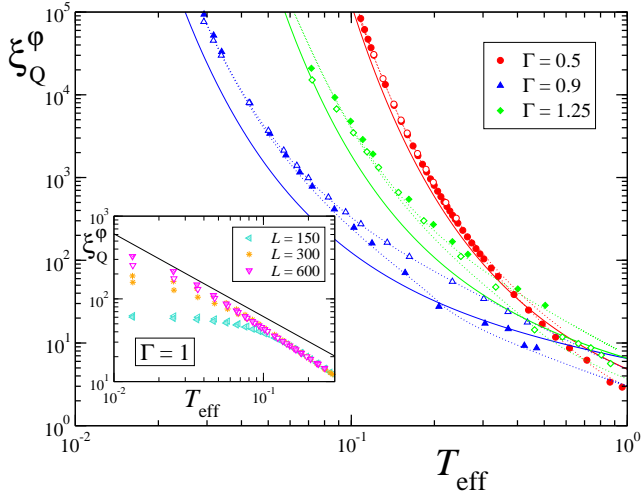}
    \caption{(color online). Correlation length as a function of
      the effective temperature, in the off-critical region for different
      values of $\Gamma$ (main panel) and at criticality (inset).
      Symbols refer to the quenched situation $\xi^\varphi_Q$;
      continuous curves denote the equilibrium values at finite
      temperatures $\xi^\varphi_T$, Eq.~\eqref{eq:coherenceR},
      while dotted lines are obtained with a semiclassical analysis
      for the system out of equilibrium, Eq.~\eqref{eq:xiSemiclass}.
      In the inset we show a finite-size scaling of data at criticality;
      the straight black curve indicates a $T_{\rm eff}^{-1}$ behavior,
      and it is plotted as a guideline.}
    \label{fig:Xi_Temp}
  \end{center}
\end{figure}

A similar peculiarity associated to the prefactor ${\tilde  K}(r)$ appears when the quench
crosses the critical point from the paramagnetic to the ferromagnetic phase. Here 
the exponential decay of spatial correlators is superimposed to oscillations 
periodically changing the sign of the correlations, ${\tilde K}(r) \approx \cos(r/r^*)$.
As shown in Fig.~\ref{fig:recurr2}, the spatial period $r^*$ of these oscillations 
varies with the distance from criticality, and eventually appears to diverge at $\Gamma_c$.
We numerically verified that $r^*$ diverges for $\Gamma \to \Gamma_c$ as
$r^*\sim 1/\sqrt{\Gamma-1}$, irrespective 
of the value of $\Gamma_0$ (in analogy to what analytically 
observed~\cite{sengupta04} for $\Gamma_0 = + \infty$).
The presence of these oscillations has been first reported 
in Ref.~[\onlinecite{sengupta04}] for a specific quench starting at $\Gamma_0=+\infty$, 
and appears to be closely analogous to a similar phenomenon observed for
linear quenches of the transverse field across the critical point~\cite{Cherng}. 
We will come back to this point in Sec.~\ref{sec:NonInt}, where we will 
show that these oscillations are an effect of the integrability of the model and that
a sufficiently strong non-integrable perturbation of the
system dynamics leads to their suppression.
Despite these oscillations, the leading exponential decay characteristic of thermal-like 
behavior is still present.
%
%
Indeed, in analogy with the phase coherence time, we find that 
the correlation length $\xi^\varphi_Q$ qualitatively follows the behavior 
of the corresponding equilibrium length $\xi^\varphi_{T}$ 
at the effective temperature $T_{\rm eff}$ determined by Eq.~\eqref{eq:Teff}.
This can be explicitly seen in Fig.~\ref{fig:Xi_Temp},
where the values $\xi^\varphi_Q$ for quantum quenches are indicated by symbols,
while the equilibrium data are plotted as continuous curves 
(see Sec.~\ref{semiclassical_out:sec} below for a discussion of the small discrepancies observed).
The dynamics at the critical point deserves a separate treatment,
since the semiclassical analysis fails there.
Again, after a finite-size scaling of numerical data, the data appear to converge at low $T_{\rm eff}$ to 
the $T_{\rm eff}^{-1}$ behavior, as predicted for the
equilibrium case at finite temperature, Eq.~\eqref{eq:xxT_crit}.
This is shown in the inset of Fig.~\ref{fig:Xi_Temp}.

\subsection{Semiclassical analysis out of equilibrium} \label{semiclassical_out:sec}

\begin{figure}[!t]
  \begin{center}
    \includegraphics[scale=0.34]{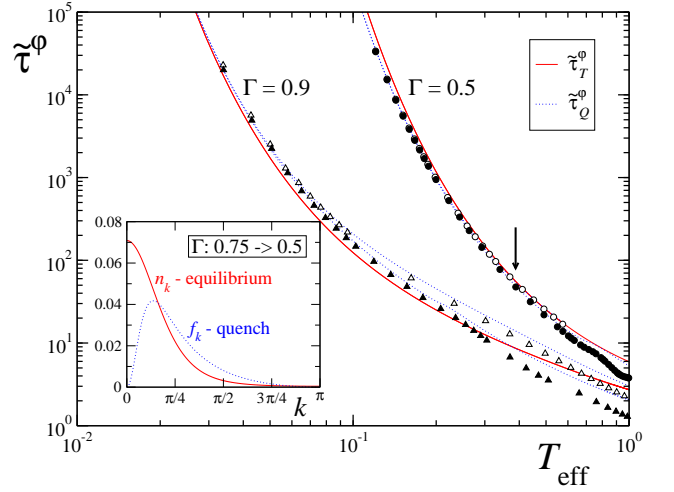}
    \caption{(color online). Phase coherence times extracted by means of
    the semiclassical argument; data are shown for two values of
    $\Gamma$ in the ferromagnetic phase.
    The continuous red curves indicate $\tilde{\tau}^\varphi_T$
    which is computed using the effective quasiparticle distribution $n_k$,
    while the dotted blue ones are for $\tilde{\tau}^\varphi_Q$ and are
    computed according to the occupation factor $f_k$.
    We also show the phase coherence times $\tau^\varphi_Q$
    evaluated   numerically  from the non-equilibrium situations
    (symbols -- same data of Fig.~\ref{fig:Tau_Temp}).
    In the inset we plot $n_k$ and $f_k$ for a particular value
    of $(\Gamma_0 = 0.75, \Gamma = 0.5)$, corresponding to the point 
    in the main panel that is indicated with an arrow.}
    \label{fig:Tau_semiclass}
  \end{center}
\end{figure}

A better agreement between numerics and the semiclassical analysis can be obtained
by modifying the analysis of Ref.~[\onlinecite{sachdev97}] for the quench case, by 
substituting the Boltzmann weight $e^{-\epsilon_k/T}$ with the occupation factor $f_k$ 
for the eigenmodes after the quench. 
In analogy with Eq.~\eqref{eq:coherenceR}, we obtain
\beq \label{eq:xiSemiclass}
\tilde{\xi}^\varphi_Q = \left( \int \! \frac{dk}{\pi} f_k \right)^{-1} \;.
\eeq
Results of the computation using this formula are shown as dashed curves 
in Fig.~\ref{fig:Xi_Temp}, and reveal a marked improvement in the quantitative 
agreement between numerics and theory.
A similar analysis applied to the autocorrelation time leads to
a generalization of Eq.~\eqref{eq:coherenceT} into:
\beq \label{eq:tauSemiclass}
\tilde{\tau}^\varphi_Q = \left( \int \! \frac{dk}{\pi} f_k \vert v_k \vert \right)^{-1} \;.
\eeq
Data displaying $\tilde{\tau}^\varphi_Q$ for two values of $\Gamma$
are shown in Fig.~\ref{fig:Tau_semiclass} (blue curves).
As one can see, the small discrepancies between $\tau^\varphi_Q$ and $\tau^\varphi_{T=T_{\rm eff}}$ 
at low temperatures that appeared in Fig.~\ref{fig:Tau_Temp} are now reconciled within numerical accuracy.

We point out that one could also evaluate the
phase coherence time by using the effective thermal quasiparticle
distribution function $n_k(T_{\rm eff})$ in Eq.~\eqref{eq:tauSemiclass},
instead of the non-equilibrium distribution function $f_k$:
$\tilde{\tau}^\varphi_T = (\frac{1}{\pi}\int  n_k(T) \vert v_k \vert dk)^{-1}$.
This gives a result that essentially coincides with the formula for
$\tau^\varphi_T$ given in Eq.~\eqref{eq:coherenceT}.
The two phase coherence times
$\tilde{\tau}^\varphi_Q$ and $\tilde{\tau}^\varphi_T$
in the low temperature regime are very close, apart from some
constant prefactor, in spite of the fact that the two quasiparticle
distribution functions $f_k$ and $n_k$ can be 
rather different, as explicitly shown in the inset of
Fig.~\ref{fig:Tau_semiclass}, for a specific value of $(\Gamma_0, \Gamma)$.

Finally, we mention that a double check of the consistency of our analysis
comes from the comparison between the effective temperature evaluated
from Eq.~\eqref{eq:Teff} and that obtained using the equality
$\tilde{\tau}^\varphi_Q = \tilde{\tau}^\varphi_T$.
In this last case, in the limit $T_{\rm eff} \ll \Delta$ 
one finds
\beq
\tilde{T}_{\rm eff} \approx \frac{\Delta}{\ln [ \Delta /(\Gamma - \Gamma_0)^2 ]} \;,
\eeq
in agreement with the cusp that emerges for $\tilde{T}_{\rm eff}$ as one approaches
the limit $\Gamma_0 \to \Gamma$ (see the inset of Fig.~(2) in Ref.~[\onlinecite{paper1}]).

\section{Non-thermal behavior of local operators} \label{sec:nonthermal}

In this section we focus on a different class of operators,
belonging to the local sector with respect to the fermionic quasiparticles.
We first consider the two-point time-dependent correlation functions
of the transverse-field magnetization $\sigma^z$, and then discuss the
behavior of the nearest-neighbor correlations of the order parameter $\sigma^x$,
which is equivalent to the density of kinks.

\subsection{The transverse-field spin correlation functions} \label{sec:ZZquench}

The two-point time-dependent correlation functions of the
transverse-field magnetization $\sigma^z$ are defined as:
\beq 
\rho^{zz} (t,r) \equiv \bra{\cal B} \sigma^z_{m+r} (t_0+t) \, \sigma^z_m (t_0) \ket{\cal B} \, ,
\eeq
in a way analogous to the order-parameter correlation functions, Eq.~\eqref{eq:XXcorr}.
These can be evaluated as sketched in Sec.~\ref{subsec:ZZtechnics}.
As for the case of $\rho^{xx}_Q(0,r)$,
we will consider a waiting time $t_0$ long enough that results are independent of it.

The correlation functions of the transverse-field operator $\sigma^z$ for the
system at equilibrium ($\Gamma_0 = \Gamma$) have been
computed in Ref.~[\onlinecite{niemeijer67}], where
it has been shown that they relax to the square of the
transverse magnetization $\langle \sigma^z \rangle$.
This is given, in the thermodynamic limit, by
\beq
   \langle \sigma^z \rangle_{\scriptscriptstyle T} =
   - \frac{1}{\pi} \int_0^\pi \cos (2 \lambda_k) \,
   \tanh \left( \frac{\epsilon_k}{2T} \right) {\rm d}k \;,
   \label{eq:MzTemp}
\eeq
with $\lambda_k = \frac{1}{2} \arctan \big( \frac{\sin k}{\cos k - \Gamma} \big)$ 
in such a way as to have $0 < \lambda_k \leq \pi$.
In contrast to the correlations of the order parameter,
at equilibrium $\rho^{zz}_T (t,r)$ exhibits a characteristic
power-law decay in time, both at zero and at finite temperature:
\beq
\vert \rho^{zz}_T (t,r) - \langle \sigma^z \rangle_{\scriptscriptstyle T}^2 \vert
\sim t^{-\alpha_T} \;,
\label{eq:ZZrelax}
\eeq
which makes it impossible to define a time-scale analogous to $\tau^{\varphi}_Q$.
The power-law exponent $\alpha_T$ 
depends on the phase of the system and on the temperature.
We calculated $\alpha_T$ by numerically fitting the curves of
$\rho^{zz}_T (t,r)$ (evaluated from exact analytic formulas~\cite{niemeijer67}) for long times $t$.
At zero temperature we found:
\beq
\alpha_0 = \left\{ \begin{array}{cl} 1 & \qquad {\rm for} \quad \Gamma < 1 \\
3/2 & \qquad {\rm for} \quad \Gamma = 1 \\ 2 & \qquad {\rm for} \quad \Gamma > 1 \;.
\end{array} \right.
\label{alpha0jump}
\eeq
On the other hand, at finite temperatures $\alpha_T = 1$, irrespective of the system phase.
More precisely, we found that the $\alpha_T = 1$ behavior always holds
for long times, while at short times and for $\Gamma \geq 1$
correlations decay with an exponent $2$,
up to a given transient time $t^*$ which decreases with the temperature.


We now consider the correlations of the transverse field $\rho^{zz}_Q(t,r)$ 
in the case of a quench and show that, contrary to the behavior of the
order parameter correlation functions~\eqref{eq:XXquench}, they do not exhibit 
thermalization.
For this purpose, we first concentrate on the asymptotic values
that are reached at long times or distances,
and then focus on the exponent of the power-law temporal decay.

As in the equilibrium case, the two-point transverse spin correlation functions
for the system out of equilibrium relax asymptotically, 
both in time and in space, to the square of the expectation value 
of the transverse spontaneous magnetization,
provided the waiting time $t_0$ after the quench is sufficiently large:
\[
\rho^{zz} (t,r) \stackrel{r,t\to\infty}{\longrightarrow} 
\langle \sigma^z \rangle_{\scriptscriptstyle Q}^2 \;.
\]
Contrary to the magnetization along the coupling direction
$\langle \sigma^x \rangle_{\scriptscriptstyle T}$,
which is rigorously zero at any finite temperature, the transverse
magnetization $\langle \sigma^z \rangle_{\scriptscriptstyle T}$
is finite even at $T>0$, see Eq.~\eqref{eq:MzTemp}.
It would then be natural to carry out an analysis on
$\langle \sigma^z \rangle_{\scriptscriptstyle Q}$,
analogous to the one performed for the relaxation dynamics
of the order parameter correlations, to check if it has anything
to do with the equilibrium transverse magnetization 
$\langle \sigma^z \rangle_{\scriptscriptstyle T_{\rm eff}}$.

\begin{figure}[!t]
  \begin{center}
    \includegraphics[scale=0.34]{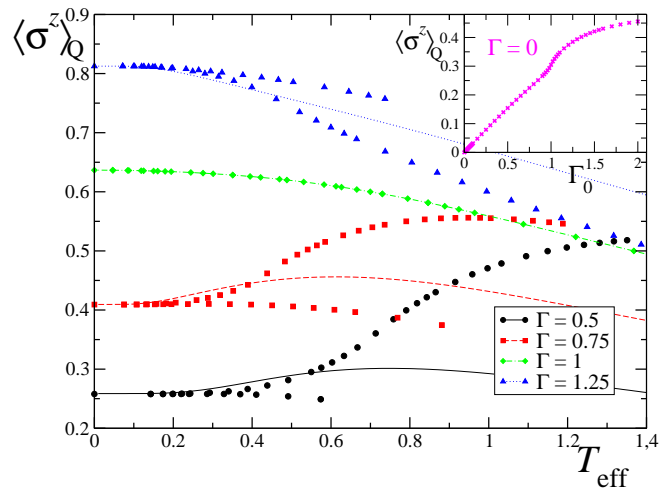}
    \caption{(color online). Asymptotic value reached by the
      transverse-field correlation functions, which is the square of
      the magnetization $\langle \sigma^z \rangle$.
      The magnetization for a quenched dynamics
      (symbols) corresponding to a given effective temperature,
      as defined in Eq.~\eqref{eq:Teff}, is compared to the corresponding
      value at equilibrium, at the same temperature (continuous lines).
      Different colors and symbols stand for the various values of $\Gamma$,
      as declared in the legend: $\Gamma=0.5$ (black circles),
      0.75 (red squares), 1 (green diamonds), 1.25 (blue triangles).
      Inset: $\langle \sigma^z \rangle_{\scriptscriptstyle Q}$ reached asymptotically
      after a quench from a given $\Gamma_{\scriptscriptstyle 0}$, to $\Gamma = 0$.
      Notice that the corresponding equilibrium value 
      $\langle \sigma^z \rangle_{\scriptscriptstyle T}$
      for $\Gamma = 0$ is rigorously zero, at any temperature.}
    \label{fig:CorrZZ_asynt}
  \end{center}
\end{figure}

We have found, however, a completely different behavior.
A signature of this fact is evident from the inset of Fig.~\ref{fig:CorrZZ_asynt},
where we plotted the value of the transverse magnetization 
$\langle \sigma^z \rangle_{\scriptscriptstyle Q}$ which is asymptotically reached 
after a quench towards $\Gamma = 0$.
As one can clearly see, contrary to the finite-temperature equilibrium
value $\langle \sigma^z \rangle_{\scriptscriptstyle T}$ which is rigorously
zero for $\Gamma = 0$ at any finite temperature $T$ [see Eq.~\eqref{eq:MzTemp}], 
the value after the quench is always non-zero, as long as $\Gamma_0 \neq 0$.
To be more quantitative, after defining an effective temperature
for the out-of-equilibrium system according to Eq.~\eqref{eq:Teff},
one discovers that 
\[
\langle \sigma^z \rangle_{\scriptscriptstyle Q} \ne 
\langle \sigma^z \rangle_{\scriptscriptstyle T_{\rm eff}} \;.
\]
This is explicitly shown in Fig.~\ref{fig:CorrZZ_asynt},
where we plot $\langle \sigma^z \rangle_{\scriptscriptstyle Q}$
as a function of $T_{\rm eff}$ (filled symbols), together with the corresponding values
$\langle \sigma^z \rangle_{\scriptscriptstyle T}$ at the
same temperature (continuous lines).
Contrary to what is observed in Figs.~\ref{fig:Tau_Temp}-\ref{fig:Xi_Temp},
here we do not find a thermal behavior,
except for the very specific case where the system is quenched towards the critical
point $\Gamma_c$ (green diamonds and line in Fig.~\ref{fig:CorrZZ_asynt}).
This can be understood in terms of the behavior of the quasi-primary operators 
of the Conformal Field Theory of the critical point~\cite{calabrese06}. 

Let us now briefly discuss the finite-time transient behavior
of $\rho^{zz}_Q (t,0)$ for the system out of equilibrium.
In a way analogous to the finite-temperature case, the transverse-field
correlations $\rho^{zz}_Q (t,0)$ relax in time as a power-law:
\[
\vert \rho^{zz}_Q (t,0) - \langle \sigma^z \rangle_{\scriptscriptstyle Q}^2 \vert
\sim t^{-\alpha_Q} \;.
\]
We found the following power-law exponents $\alpha_Q$, depending on the
system phases before and after the quench:
\begin{center}
\begin{tabular}{c|c|c|c}
$\alpha_{\scriptscriptstyle Q}
       $     & \, $\Gamma_0 < 1$ \, & \, $\Gamma_0 = 1$ \, & $\Gamma_0 > 1$ \\ \hline
$\Gamma < 1$ & 1                    & 1                    & 2            \\
$\Gamma = 1$ & 2                  & 3/2                  & 2            \\
$\Gamma > 1$ & 1                    & 1                    & 2
\end{tabular}
\end{center}
Since for the system at equilibrium and at any finite temperature
we have $\alpha_T = 1$, this means that, in general, even the finite-time
power-law behavior is a non-thermal one.
The jump in the values of $\alpha_0$ in Eq.~\eqref{alpha0jump} 
and of $\alpha_{\scriptscriptstyle Q}$ in the table, depending on the different system 
phases before and after the quench, can be seen as due to lattice effects.

\subsection{The density of kinks} \label{sec:kinks}

As a further confirmation of the non-thermal behavior of local
operators, we concentrate on the density of kinks $\rho^{\cal N}$.
In particular we compare the asymptotic value reached after a quench 
$\langle \rho^{\cal N} \rangle_{\scriptscriptstyle Q}$ with the thermal behavior at equilibrium.
In the thermodynamic limit this is given by~\cite{lieb61,pfeuty70}:
\beq
\langle \rho^{\cal N} \rangle_{\scriptscriptstyle T} = \int_0^\pi \frac{dk}{2 \pi} \left[ 1- 
\frac{1 + \Gamma \cos k}{\sqrt{1+\Gamma^2 + 2 \Gamma \cos k}}
\big( 1 - 2 n_k(T) \big)  \right] \,.
\label{eq:Kink_T}
\eeq
On the other hand, the out-of-equilibrium behavior of 
$\langle \rho^{\cal N} \rangle_{\scriptscriptstyle Q} \equiv \lim_{t \to \infty} \rho^{\cal N}(t)$
after a quench is retrieved from Eq.~\eqref{eq:KINKcomput} and
by taking the asymptotic long-time limit for $L \to \infty$.

\begin{figure}[!t]
  \begin{center}
    \includegraphics[scale=0.33]{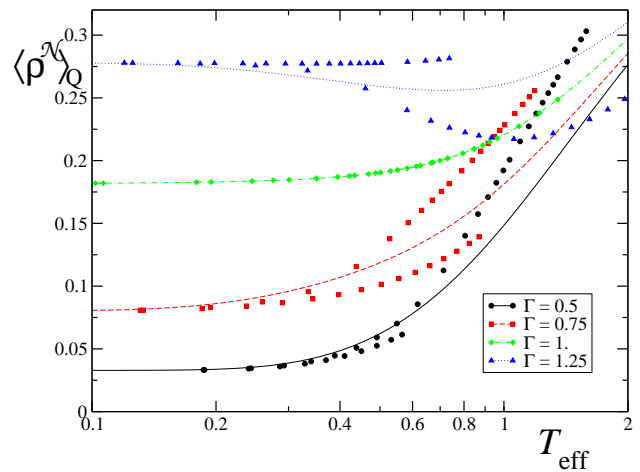}
    \caption{(color online). Asymptotic value of the density
      of kinks as a function of the effective temperature $T_{\rm eff}$.
      Different colors stand for various
      values of $\Gamma$, as shown in the inset.
      Symbols denote the density of kinks $\langle \rho^{\cal N} \rangle_{\scriptscriptstyle Q}$ 
      after a quench; to each value of $\Gamma_0$ corresponds a different
      initial state, therefore a different effective temperature $T_{\rm eff}$.
      Straight lines denote the finite-temperature equilibrium values
      $\langle \rho^{\cal N} \rangle_{\scriptscriptstyle T = T_{\rm eff}}$.}
    \label{fig:Asyng_Temp}
  \end{center}
\end{figure}

The comparison between the values of $\langle \rho^{\cal N} \rangle_{\scriptscriptstyle Q}$ 
after a quench and $\langle \rho^{\cal N} \rangle_{\scriptscriptstyle T_{\rm eff}}$ at equilibrium, 
where $T_{\rm eff}$ for the corresponding out-of-equilibrium system is obtained 
from Eq.~\eqref{eq:Teff}, is presented in Fig.~\ref{fig:Asyng_Temp}.
%
%
As one can see, outside criticality the two quantities 
are evidently not related and behave in different ways.
On the other hand, the results for quenches ending at the critical point $\Gamma_c=1$
(diamonds) perfectly follow the dotted-dashed equilibrium line (green data),
so that the density of kinks after a quench to the critical point
is univocally determined by the effective temperature $T_{\rm eff}$.
Remarkably such behavior is not found for a non-critical dynamics,
where fine details of the initial condition seem to be important.
As explained above, we frame the effective thermalization only at 
criticality in terms of the behavior of the quasi-primary operators 
in conformal field theory~\cite{calabrese06}.

Incidentally, the thermal behavior of $\langle \rho^{\cal N} \rangle_Q$
at criticality can be recovered analytically.
Indeed, in that case the density of kinks at equilibrium
with an effective temperature $T_{\rm eff}$ is given by
substituting $\Gamma=\Gamma_c=1$ in Eq.~\eqref{eq:Kink_T}:
\barr
 \left. \langle \rho^{\cal N} \rangle_{\scriptscriptstyle T_{\rm eff}} \right|_{\rm cr} 
 &=& \int_0^{\pi} \frac{dk}{2 \pi}
 \left[ 1- \frac{1}{4}\epsilon_k^{\Gamma_c} \big[ 1 - 2 n_k(T_{\rm eff}) \big]  \right] \nonumber \\
 &=& \frac{1}{2} + \frac{ \langle \Ham(\Gamma) \rangle_{T_{\rm eff}} }{4} \;,
\label{eq:Teff_G1}
\earr
where $\epsilon_k^{\Gamma_c} = 2\sqrt{2 - 2\cos k}$ is the energy of the ${\cal A}_k$
quasiparticle at $\Gamma_c=1$, and the second equality 
follows from Eq.~\eqref{eq:Teff} with $n_k(T) = n_{-k}(T)$.
The energy $\langle \Ham(\Gamma) \rangle_{T_{\rm eff}}$ of the system after
the quench is obtained directly from Eq.~\eqref{eq:encalc}, 
by evaluating $\langle {\cal B} \vert {\cal A}^\dagger_k {\cal A}_k \vert {\cal B} \rangle$, 
and is given by
\beq
 \langle \Ham(\Gamma) \rangle_{T_{\rm eff}} = - \int_0^{\pi} \frac{dk}{\pi}
  \frac{2(1 + \Gamma_0) (1 + \cos k)}{\epsilon_k^{\Gamma_0}} \;.
\eeq
On the other hand, the value of $\langle \rho^{\cal N} \rangle_{\scriptscriptstyle Q}$ 
for a quench at criticality is extracted from Eq.~\eqref{eq:KINKcomput} for $t \to \infty$: 
\barr
\left. \langle \rho^{\cal N} \rangle_{\scriptscriptstyle Q} \right|_{\rm cr} = 
\int_0^{\pi} \frac{dk}{2\pi}  \bigg\{ 1 - 
 2\frac{1+\Gamma_0 \cos k}{\epsilon_k^{\Gamma_0}} 
  - 8\frac{(\Gamma_0 - 1) \sin^2 k} {(\epsilon_k^{\Gamma_c})^2 \, \epsilon_k^{\Gamma_0}} \bigg\}  \nonumber
\earr
which, after simple algebra, can be shown to reduce to Eq.~\eqref{eq:Teff_G1}, 
hence obeying the rigorous equality 
\beq
\left. \langle \rho^{\cal N} \rangle_{\scriptscriptstyle Q} \right|_{\rm cr} 
= \left. \langle \rho^{\cal N} \rangle_{\scriptscriptstyle T_{\rm eff}} \right|_{\rm cr} \;.
\eeq

The very similar behavior of $\langle \sigma^z \rangle$ and $\langle \rho^{\cal N} \rangle$
after a quench follows from the structure of the Ising Hamiltonian 
in Eq.~\eqref{eq:model}, which is a sum of the kink-operator 
term plus a transverse magnetization term, and the fact that the
total energy is a conserved quantity in the quench process.

\subsection{Description in terms of a generalized Gibbs ensemble} \label{subsec:Gibbs}

The non-thermal behavior of the long-time asymptotic values for the transverse
magnetization $\langle \sigma^z \rangle_{\scriptscriptstyle Q}$ 
and for the density of kinks $\langle \rho^{\cal N} \rangle_{\scriptscriptstyle Q}$ 
outside criticality can be reconciled
in terms of a generalized Gibbs ensemble $\rho_{\mathcal G}$~\cite{rigol07},
which takes into account all the non-trivial constants of motion that
typically prevent integrable systems from having an ergodic behavior:
\beq
\rho_{\mathcal G} = \frac{1}{Z_{\mathcal G}} \exp \Big( -\sum_k \beta_k \epsilon_k^{\Gamma} n_k \Big) \,,
\label{eq:Gibbs}
\eeq
where $Z_{\mathcal G} = {\rm Tr} [e^{-\sum_k \beta_k \epsilon_k^{\Gamma} n_k}]$
is the partition function, and $n_k = {\cal A}^\dagger_k {\cal A}_k$ is the number operator of the 
Bogoliubov quasiparticle with energy $\epsilon_k^{\Gamma}$.
In Eq.~\eqref{eq:Gibbs} the operators $\left\{ n_k \right\}$ assume the role
of a full set of constants of motion, thus keeping track of the details 
of the initial condition. 
The fictitious inverse temperature $\beta_k$ is taken to depend on mode $k$ of the system,
and is a Lagrange multiplier that can be calculated from 
an equation for the expectation value of the corresponding integral of motion $n_k$
for the evolving system after the quench $\Ham (\Gamma)$, on the boundary state $\ket{\cal B}$:
\beq
\langle n_k\rangle_{\mathcal G} = (1+e^{{\beta_k \epsilon_k^{\Gamma}}})^{-1}
= \bra{\cal B} n_k \ket{\cal B} =f_k \,.
\label{eq:betaFict}
\eeq
After expressing the number operator $n_k$ in terms of the creation and
annihilation operators for the initial Hamiltonian $\Ham(\Gamma_0)$, one arrives at
the following expression for the r.h.s. of Eq.~\eqref{eq:betaFict}:
\beq
f_k = \bra{\phi(\Gamma_0)} n_k \ket{\phi(\Gamma_0)} = \frac{1}{2} - \frac{a_k^\Gamma a_k^{\Gamma_0} + b_k^2}
{2 \epsilon_k^\Gamma \epsilon_k^{\Gamma_0}} \, ,
\eeq
which, together with Eq.~\eqref{eq:betaFict}, provides an explicit formula for the 
$\beta_k$'s~\footnote{The fictitious temperature $\beta_k$ defined in Eq.~\eqref{eq:betaFict}
depends on mode $k$ also at criticality, since it is obtained by equating 
the contribution to the total energy due to a single-mode population.
Therefore in principle there is no connection with the effective thermodynamic temperature
$\beta$ of Eq.~\eqref{eq:Teff}. This is the case even at $\Gamma_c$, where the system exhibits 
a true thermal behavior in the sense of standard statistical ensembles.}.

Using such obtained $\beta_k$, one finds that the asymptotic values 
of both the transverse magnetization and the density of kinks after a quench are identical 
to the corresponding expectation values in the same generalized Gibbs ensemble, 
irrespective of the initial and the final transverse fields $\Gamma_0 , \Gamma$:
\beq
\langle \sigma^z \rangle_{\scriptscriptstyle Q} = \langle{\sigma^z}\rangle_{\mathcal G}
\qquad \langle \rho^{\cal N} \rangle_{\scriptscriptstyle Q} = \langle{\cal N}\rangle_{\mathcal G} \, .
\eeq
This shows that the Gibbs distribution~\eqref{eq:Gibbs},
with the same fictitious temperatures $\beta_k$ defined in Eq.~\eqref{eq:betaFict},
is able to catch the non-thermal behavior of both $\sigma^z$ and ${\cal N}$ 
after the relaxation outside criticality.
As pointed out before, the two quantities necessarily have to behave in the same way,
due to the energy conservation after the quench 
and to the particular structure of the Hamiltonian~\eqref{eq:model}.

\section{Sensitivity to the breaking of integrability} \label{sec:NonInt}

In this section we briefly address the problem of how a non-integrable perturbation
of the Ising model~\eqref{eq:model} can affect the relaxation to the steady state.
We do not intend to give here a fully comprehensive study of the effects of
integrability breaking: more modestly, the purpose of this section is to provide 
evidence that some features of the asymptotic equal-time correlators, 
in particular the spatial oscillations after a quench, 
disappear as soon as the non-integrable perturbation is increased.
This sheds light on the scenario we proposed throughout the paper: strong effects 
on the local observables (which do not thermalize by themselves, in integrable systems) 
are expected, while the qualitative behavior of non-local ones (which, on the contrary,
are already ``thermal'') should not change.

One way to break integrability of the Ising model is to add
a next-to-nearest-neighbor coupling of strength $J_2$:
\beq
   \Ham_2(\Gamma) = -J \sum_{j=1}^{L} \left[ \sigma^x_j \sigma^x_{j+1}
   + \Gamma \sigma^z_j \right] -J_2 \sum_{j=1}^{L}  \sigma^x_j \sigma^x_{j+2} \;.
   \label{eq:model2}
\eeq
For such non-integrable system, the standard analytic tools based 
on the Jordan-Wigner transformation fail, therefore one generally has to resort
to a fully numerical treatment. Here we employ an exact diagonalization study
of model~\eqref{eq:model2}; unfortunately, this severely limits the actual sizes
of the systems under consideration to $\sim O(10)$ spins.

\begin{figure}[!t]
  \begin{center}
    \includegraphics[scale=0.33]{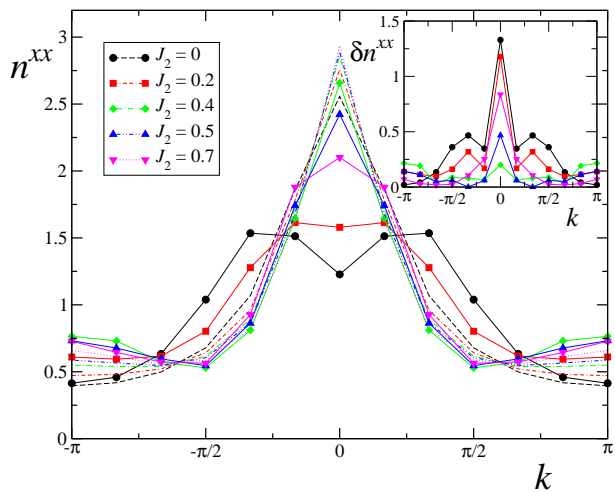}
    \caption{(color online). Fourier transform of the equal-time order-parameter 
      correlation function for the Ising model in presence of
      a non-integrable perturbation $J_2$.
      Symbols connected by lines denote $n^{xx}_Q(k)$ after a quench from $\Gamma_0 = 5$
      to $\Gamma = 0.4$, according to the diagonal ensemble prediction.
      Dashed/dotted lines stand for the corresponding expectation values in
      the canonical ensemble $n^{xx}_{T_{\rm eff}}(k)$ at the corresponding 
      effective temperature $T_{\rm eff}$.
      Inset: absolute differences $\delta n^{xx}(k)$ between the diagonal 
      and the canonical ensemble predictions.
      Data are for a chain with $L=12$ sites.}
    \label{fig:IntBreak_FM}
  \end{center}
\end{figure}

We discuss the effective ``generic'' thermalization induced
by integrability breaking, by analyzing the changes in the space asymptotics
of the order parameter correlators, after a quench in the transverse-field 
strength $\Gamma$: as discussed in Sec.~\ref{subsec:XXspacedep},
for quenches crossing the Ising quantum critical point from the paramagnetic to the ferromagnetic side,
non-thermal sinusoidal oscillations in real space naturally emerge~\cite{sengupta04,Cherng}.
Below we numerically show that a non-integrable perturbation leads to a 
suppression of them, thus driving the system towards a completely thermal-like behavior,
where only the exponential decay in space of correlations is present.
Instead of considering the correlations $\rho^{xx}_Q(0,r)$ in real space,
here we work in the momentum space, where the effects of the oscillations clearly 
emerge also for very small sizes. We define
\beq
n^{xx}(k) \equiv \frac{1}{L} \sum_{r=1}^L e^{i r k/L} \rho^{xx}_Q(0,r) \,.
\eeq
In presence of a pure exponential decay in space of correlations $\rho^{xx}_Q$,
a typical Lorentzian curve for $n^{xx}(k)$ emerges.
The superposition of a simple sinusoidal modulation induces the formation
of two humps in $k$-space, as it can be seen from the black continuous curve (circles) 
in Fig.~\ref{fig:IntBreak_FM}, for $J_2 = 0$.

In Fig.~\ref{fig:IntBreak_FM} we numerically compare the expectation value 
$n^{xx}_{T_{\rm eff}}(k)$ in the canonical ensemble at an effective temperature
$T_{\rm eff}$ as given by Eq.~\eqref{eq:Teff}, with the asymptotic value
after the quench $n^{xx}_Q(k)$, calculated from the diagonal ensemble~\cite{rigol09}:
\beq
n^{xx}_Q(k) \equiv \lim_{t \to \infty} \bra{\phi_t} n^{xx}(k) \ket{\phi_t}
= \sum_i \vert c_i \vert^2 \bra{\varphi_i} n^{xx}(k) \ket{\varphi_i} \,,
\eeq
where $\ket{\phi_t}$ is the state at time $t$ after the quench
$\ket{\phi_t} = e^{- i \Ham_2(\Gamma) t} \ket{\cal B}$,
$\ket{\varphi_i}$ are the eigenstates of the Hamiltonian after the quench $\Ham_2(\Gamma)$,
and $c_i = \langle {\cal B} \ket{\varphi_i}$.
Humps are present in the out-of-equilibrium system for very small values of $J_2$,
while they gradually disappear when increasing the perturbation;
on the contrary, they are completely absent at equilibrium at finite temperature,
thus clearly suggesting that they are typical non-thermal features.
A clear evidence of the thermalization induced by non-integrability is given
by the difference $\delta n^{xx}(k) \equiv \vert n^{xx}_{T_{\rm eff}}(k) -n^{xx}_Q(k) \vert$
as a function of $J_2$. As shown in the inset of Fig.~\ref{fig:IntBreak_FM},
$\delta n^{xx}(k)$ is decreasing with $J_2$, until this becomes large and the 
system turns out to be again integrable.

\section{Conclusions} \label{sec:concl}

In summary, we have analyzed the temporal relaxation of some 
observables after a sudden quantum quench in a completely integrable 
one-dimensional spin model exhibiting a quantum phase transition.
In particular, we focused on the spin-$1/2$ quantum Ising chain, where we 
quenched the transverse magnetic field and studied the space and time asymptotics
of two-point spin correlation functions corresponding to the spin operators 
along the longitudinal and the transverse direction, and the density of kinks.

We found that, despite the complete integrability of the Ising model,
some observables exhibit thermal behavior, while others do not.
The thermal behavior is typical for operators which are non-local
with respect to the fermion quasiparticles that diagonalize the model. 
The spin operator along the coupling direction, which is the order
parameter, is of such type. In this paper we have 
explicitly demonstrated that correlation functions of the order parameter
drop exponentially to zero and thermalize to an effective temperature
which is ruled by the energy of the initial state after the quench.
On the other hand, as examples of operators which are local
in the quasiparticles, we considered the spin operator
along the transverse field direction, and the two-point nearest-neighbor
correlator of the spin operator along the coupling direction.
We showed that such two-point correlations present non-thermal
power-law decay to some residual value, which is not
related to its thermal counter-part at the associated effective temperature,
except in the very specific case of a quench towards the critical point.

It is worth stressing that after breaking the integrability of the model by adding 
an extra operator to the Hamiltonian [a next-to-nearest neighbor coupling, see Eq.~\eqref{eq:model2}], 
spurious non-thermal effects of non-local operators disappear, 
while the leading thermal behavior does not qualitatively change. 
In particular, we have provided numerical evidence for the suppression of non-thermal sinusoidal 
oscillations in real space of the order parameter correlators, 
after a quench from the paramagnetic to the ferromagnetic phase.

We point out that the relaxation dynamics of a many body system
following a quantum quench has been considered also in other similar contexts.
In Ref.~[\onlinecite{barmettler09}] the behavior of the staggered transverse
magnetic moment in a XXZ spin chain initialized in the perfect antiferromagnetic
state was analyzed: the staggered moment always decays exponentially, similar to 
our findings on the order parameter correlations in the Ising model. 
In some cases, depending on the value of the anisotropy parameter,
oscillations in the dynamics of the magnetization, which resemble the ones we reported 
for quenches from the paramagnet to the Ising ferromagnet~\cite{sengupta04,Cherng},
were also found.
The time evolution of an initial state equivalent to the N\'eel state
has been also studied for the Bose-Hubbard model~\cite{cramer08B}.
In this case, oscillations of local observables such as the local density
or nearest-neighbor correlators, seem to decay algebraically for all values of the
interaction, while no crossover to a non oscillatory regime has been found.
The differences between the two cases have been ascribed to the role
played by the crossing of equilibrium critical points. 

In light of all these results, it would be clearly interesting to investigate other quantum integrable models, 
in order to see whether thermalization behavior occurs in correlation 
functions of operators that couple to infinitely many quasi-particle 
states of the theory, while it is absent in correlation functions of 
operators with a finite number of matrix elements.

\acknowledgments
We acknowledge interesting discussions with Natan Andrei,
Pasquale Calabrese, Elena Canovi, Rosario Fazio, Davide Fioretto, Andrea Gambassi, Vladimir Gritsev, Vladimir Kravtsov, 
Markus M\"uller, Alexander Nersesyan, Anatoli Polkovnikov. G.M. acknowledges the grants INSTANS
(from ESF) and 2007JHLPEZ (from MIUR).
S.S. is supported by Grant-in-Aid for Scientific Research from MEXT, Japan.


\end{document}